\documentclass[useAMS,usenatbib,usegraphicx,fleqn]{mn2e}

\usepackage{amsmath}
\usepackage{url}
\usepackage{times}

\usepackage{amssymb}  
\def \aap{A\&A}
\def \aaps{A\&AS}

\def \aj{AJ}

\def \apj{ApJ}
\def \apjl{ApJ}
\def \apjs{ApJS}
\def \araa{ARA\&A}

\def \mnras{MNRAS}

\def \nat{Nat}

\def \pasp{PASP}

\usepackage{color}
\usepackage{graphicx}

\newcommand{\AlI}{\hbox{{\rm Al}{\sc \,i}}}

\newcommand{\CI}{\hbox{{\rm C}{\sc \,i}}}

\newcommand{\CII}{\hbox{{\rm C}{\sc \,ii}}}
\newcommand{\CIIs}{\hbox{{\rm C}{\sc \,ii}$^*$}}
\newcommand{\CIII}{\hbox{{\rm C}{\sc \,iii}}}
\newcommand{\CIV}{\hbox{{\rm C}{\sc \,iv}}}
\newcommand{\NV}{\hbox{{\rm N}{\sc \,v}}}
\newcommand{\CaI}{\hbox{{\rm Ca}{\sc \,i}}}
\newcommand{\CaII}{\hbox{{\rm Ca}{\sc \,ii}}}

\newcommand{\FeI}{\hbox{{\rm Fe}{\sc \,i}}}
\newcommand{\FeII}{\hbox{{\rm Fe}{\sc \,ii}}}
\newcommand{\Htwo}{\hbox{{\rm H}$_2$}}
\newcommand{\HI}{\hbox{{\rm H}{\sc \,i}}}

\newcommand{\MgI}{\hbox{{\rm Mg}{\sc \,i}}}
\newcommand{\MgII}{\hbox{{\rm Mg}{\sc \,ii}}}
\newcommand{\NI}{\hbox{{\rm N}{\sc \,i}}}
\newcommand{\NII}{\hbox{{\rm N}{\sc \,ii}}}
\newcommand{\NaI}{\hbox{{\rm Na}{\sc \,i}}}

\newcommand{\OI}{\hbox{{\rm O}{\sc \,i}}}
\newcommand{\OII}{\hbox{{\rm O}{\sc \,ii}}}

\newcommand{\OVI}{\hbox{{\rm O}{\sc \,vi}}}

\newcommand{\SiII}{\hbox{{\rm Si}{\sc \,ii}}}
\newcommand{\SiIII}{\hbox{{\rm Si}{\sc \,iii}}}
\newcommand{\SiIV}{\hbox{{\rm Si}{\sc \,iv}}}
\newcommand{\ZnII}{\hbox{{\rm Zn}{\sc \,ii}}}
\newcommand{\ZnI}{\hbox{{\rm Zn}{\sc \,i}}}
\newcommand{\TiII}{\hbox{{\rm Ti}{\sc \,ii}}}
\newcommand{\MnII}{\hbox{{\rm Mn}{\sc \,ii}}}

\newcommand{\nH}{n_{\rm H}}

\newcommand{\lya}{\mbox{Ly$\alpha$}}
\newcommand{\lyb}{\mbox{Ly$\beta$}}

\newcommand{\kms}{\mbox{km~s$^{-1}$}}
\newcommand{\cmm}{\mbox{cm$^{-2}$}}
\newcommand{\cmmm}{\mbox{cm$^{-3}$}}

\newcommand{\msunyr}{\mbox{M$_{\sun}$~yr$^{-1}$}}

\newcommand{\NHI}{\hbox{$N_\textsc{H\,i}$}}
\newcommand{\NHII}{\hbox{$N_\textsc{H\,ii}$}}
\newcommand{\NHtwo}{\hbox{$N_{\textrm{H}_2}$}}
\newcommand{\fHtwo}{\hbox{$f_{\textrm{H}_2}$}}

\title[A high molecular fraction in a sub-DLA at $z=0.56$]{A high
  molecular fraction in a sub-damped absorber at
  $\bmath{z=0.56}$\thanks{All of the data and much of code used in
    this paper are available at
    \protect\url{https://github.com/nhmc/H2}.}}

\author[N. Crighton et al.]{Neil H. M. Crighton$^{1,
    2}$\thanks{E-mail: neilcrighton@gmail.com}, Jill Bechtold$^{3}$,
  Robert F. Carswell$^{4}$, Romeel Dav\'e$^3$, \newauthor Craig
  B. Foltz$^{5}$, Buell T. Jannuzi$^{3, 6}$, Simon L. Morris$^{2}$,
  John M. O'Meara$^7$, \newauthor J. Xavier Prochaska$^8$, Joop
  Schaye$^9$ and Nicolas Tejos$^{2}$
\\
\\
  $^1$Max-Planck-Institut f{\"u}r Astronomie, K{\"o}nigstuhl 17, D-69117 Heidelberg, Germany. \\
  $^2$Department of Physics, University of Durham, South Road, Durham DH1 3LE, U.K. \\
  $^3$Department of Astronomy, University of Arizona, Tucson, AZ 85721, USA.\\
  $^4$Institute of Astronomy, Madingley Road, Cambridge, CB3 0HA, U.K.\\
  $^5$Division of Astronomical Sciences, National Science Foundation, 4201 Wilson Blvd. Arlington, VA 22230.\\
  $^6$National Optical Astronomy Observatory, 950 N. Cherry Ave., Tucson, Arizona, 85719, USA.\\
  $^7$Department of Chemistry and Physics, Saint Michael’s College. One Winooski Park, Colchester, VT 05439. \\
  $^8$Department of Astronomy and Astrophysics \& UCO/Lick Observatory, University of California, 1156 High Street, Santa Cruz, CA 95064, USA.\\
  $^9$Leiden Observatory, Leiden University, P.O. Box 9513, 2300 RA Leiden, the Netherlands.
}

\begin{document}

\date{Accepted xxxx. Received xxxx; in original form xxxx}

\pagerange{\pageref{firstpage}--\pageref{lastpage}} \pubyear{xxxx}

\maketitle

\label{firstpage}

\begin{abstract}

Measuring rest-frame ultraviolet rotational transitions from the Lyman
and Werner bands in absorption against a bright background continuum
is one of the few ways to directly measure molecular hydrogen
(H$_2$). Here we report the detection of absorption from H$_2$ at
$z=0.56$ in a sub-damped \lya\ system with neutral hydrogen column
density \NHI\,$=10^{19.5 \pm 0.2}$\,\cmm.  This is the first H$_2$
system analysed at a redshift $<1.5$ beyond the Milky Way halo. It has
a surprisingly high molecular fraction: log$_{10}$\,\fHtwo\,$> -1.93
\pm 0.36$ based on modelling the line profiles, with a robust
model-independent lower limit of \fHtwo\,$>10^{-3}$. This is higher
than \fHtwo\ values seen along sightlines with similar \NHI\ through
the Milky Way disk and the Magellanic clouds.  The metallicity of the
absorber is $0.19^{+0.21}_{-0.10}$ solar, with a dust-to-gas ratio $<
0.36$ of the value in the solar neighbourhood. Absorption from
associated low-ionisation metal transitions such as \OI\ and \FeII\ is
observed in addition to \OVI. Using \textsc{Cloudy} models we show
that there are three phases present; a $\sim 100$\,K phase giving rise
to H$_2$, a $\sim 10^4$\,K phase where most of the low-ionisation
metal absorption is produced; and a hotter phase associated with
\OVI. Based on similarities to high velocity clouds in the Milky Way
halo showing H$_2$, and the presence of two nearby galaxy candidates
with impact parameters of $\sim10$\,kpc, we suggest that the absorber
may be produced by a tidally-stripped structure similar to the
Magellanic Stream.

\end{abstract}


\section{Introduction}

Molecular hydrogen (H$_2$) is the most abundant molecule in the
universe and is closely linked to star formation via the star
formation surface density rate -- molecular gas surface density
relation \citep{Bigiel08}.  Measuring rest-frame UV rotational
transitions from the Lyman and Werner bands in absorption against a
bright background continuum is one of the few ways to directly measure
H$_2$ \citep[see][for example]{Draine11}. This technique probes
diffuse gas with molecular fractions, \fHtwo, of $\sim 10^{-6}$ to
$\sim 0.1$ -- denser molecular clouds are both dusty, and thus likely
to extinguish UV light from a background source, and compact, such
that there is a low probability of intersection with a sightline to a
background light source \citep[][]{Hirashita03, Zwaan06}. However, the
lower molecular fraction systems that are detected give valuable
insights into the environments and physical mechanisms necessary for
the formation of H$_2$. With this technique we can measure the
physical properties of cool, dense gas over a large fraction of the
age of the Universe, from the interstellar medium in the solar
neighbourhood to proto-galaxies a few Gyr after the big bang.

Since the initial detection towards the UV bright star $\xi$ Persei
\citep{Carruthers70}, a large sample of sightlines exhibiting H$_2$ in
absorption from the Milky Way and its halo has been assembled. These
observations have characterised H$_2$ in diffuse molecular gas in the
Milky Way plane \citep{Savage77}, the Magellanic clouds
\citep{Tumlinson02, Welty12}, high latitude sightlines out of the
Milky Way plane \citep{Gillmon06, Wakker06}, in intermediate and high
velocity clouds \citep[IVCs and HVCs,][]{Richter03b,Richter99}, and in
the Magellanic Stream \citep{Sembach01, Richter01b}. A physical
picture where H$_2$ formation occurs predominantly on the surface of
dust grains \citep{Shull82} in clouds with total densities of $n\sim
10-100$\,\cmmm\ illuminated by the local UV radiation has been
successful in reproducing both the observed H$_2$ rotational
population levels and molecular fractions in the Milky Way (for
example \citealt{Spitzer74}; \citeauthor{Jura75b}
\citeyear{Jura75a},b) and the Magellanic clouds \citep{Tumlinson02}.

H$_2$ has also been measured at redshifts $1.5 - 4.5$, corresponding
to lookback times of $\sim 9 - 12$\ Gyr, in damped
\lya\ (\NHI\ $>10^{20.3}$\,\cmm, DLA) and sub-damped
\lya\ ($10^{19}$\,\cmm\,$\lesssim$~\NHI~$\lesssim 10^{20.3}$\,\cmm,
sub-DLA) absorption systems seen towards bright background QSOs. In
this redshift range, absorption features from \HI\ and sometimes H$_2$
are redshifted into the optical range, making them relatively easy to
detect with large ground-based telescopes. The first unambiguous
detection in a redshifted absorber was made by \citeauthor{Foltz88}
\citep[1988, see also][]{Levshakov85}, and since then at least $16$
further such systems have been discovered \citep[for example][]{Ge97,
  Ge01, Levshakov02, Cui05, Ledoux06, Noterdaeme07}. Approximately
$10$ per cent of DLAs have a molecular fraction \fHtwo\,$>10^{-4.5}$,
and these tend to be more metal rich and dustier \citep{Ledoux03}, and
have higher velocity widths \citep{Noterdaeme08} than DLAs without
detectable H$_2$. Several physical diagnostics are available to
measure the properties of the H$_2$ absorbing gas. Some H$_2$ systems
also show absorption from the CO molecule, revealing the presence of a
cold, dense core of gas with excitation temperatures consistent with
those expected from the cosmic microwave background
\citep[for example][]{Srianand08, Noterdaeme09}. The H$_2$ rotational level
populations and \CI\ fine structure transitions can also be used to
measure particle densities. They are generally found to be similar to
those measured along local sightlines in the Milky Way ($\sim
10-100$\,\cmmm), but the ambient UV field, gas temperatures and gas
pressures tend to be higher \citep{Hirashita05, Srianand05}.

No studies currently exist of H$_2$ at lower redshifts, $z < 1.5$,
outside the Milky Way halo. Until recently, the low number of DLA and
sub-DLA systems known at low redshifts, together with the smaller
light gathering power of spaced-based UV telescopes compared to large
aperture ground-based optical telescopes, have made observing the
Lyman-Werner bands in this redshift range impractical. However, with
the availability of the far-UV sensitive Cosmic Origins Spectrograph
(COS) on the Hubble Space Telescope (HST), molecular absorption can
now be effectively detected for $0.1 \lesssim z \lesssim 0.8$.

In this paper we report the serendipitous detection of H$_2$ in a
sub-DLA at $z=0.56$, the first such system analysed at a redshift
below $1.5$ beyond the Milky Way halo. It has a high molecular
fraction given the total cloud neutral hydrogen column density, and we
show that the associated metal absorption features seen require the
presence of three phases: a cold $T\sim 100$\,K phase analogous to the
cold neutral medium observed in the Milky Way's interstellar medium
(ISM); a partially-ionised $T\sim 10^4$\,K phase, similar to the warm
neutral medium in the ISM; and a warmer, probably collisionally
ionised phase. Based on the cloud properties we argue the absorber is
likely caused by a tidally-stripped absorbing structure similar to the
Magellanic Stream embedded in a warm halo $\sim 10$\,kpc from a nearby
galaxy.

The layout of the paper is as follows. Section 2 describes the data
used; Section 3 describes how we identified lines and measured the
absorption line properties; and Section 4 describes the properties of
the H$_2$ absorption and the sub-DLA. We compare to theoretical models
and discuss our results in Section 5, and summarise the main results
of the paper in Section 6. When not explicitly shown logarithms are to
base 10, and we use a 7-year WMAP cosmology
\citep[$H_0=70.4$~\kms$\,$Mpc$^{-1}$, $\Omega_\mathrm{M}=0.272$,
  $\Omega_\Lambda=0.728$;][]{Komatsu11} where necessary. We use
transition wavelengths and oscillator strengths given by
\citet{Morton76, Morton03} and \citet{Verner94}, and H$_2$ transition
wavelengths from \citet{Bailly10}.

\section{Data}

Transitions from the sub-DLA are measured in absorption against the
continuum from the background QSO, Q~0107$-$0232, at $z_{\rm qso}=0.728$
(see Table~\ref{t_qso}). This was discovered by the Large Bright
Quasar survey \citep{Hewett95} and is one of a group of three bright
QSOs with small angular separations on the sky. Spectra of these QSOs
taken using the Faint Object Spectrograph (FOS) on the HST have been
used to measure correlations in neutral hydrogen absorption
\citep{Young01, Petry06} and in absorption with galaxy positions
\citep{Crighton10} across the three sightlines.

Here we present higher resolution far UV spectra of Q~0107$-$0232
taken with the Cosmic Origins Spectrograph on the HST, and an optical
spectrum taken with the High Resolution Echelle Spectrograph (HIRES)
on Keck I. In our analysis we also make use of $K$ band imaging of the
QSO and archival UV FOS spectra. The FOS spectra were originally
published by \citet{Young01}. We employ the combined spectrum used by
\citet{Crighton10}, covering a wavelength range of
$1572-2311\,$\AA\ at a typical signal to noise (S/N) of $31$ per
$4\,$\AA\ resolution full width at half maximum intensity (FWHM).

\begin{table}
\addtolength{\tabcolsep}{-1.5pt}
\begin{center}
\begin{tabular}{cccccc}
Name & R.A. (J2000) & Dec. (J2000) & $z_{\rm em}$ & $R$-mag  \\
\hline
Q~0107$-$0232 & $01^\mathrm{h}10^\mathrm{m}14.43^\mathrm{s}$ 
& $-02\degr16\arcmin57.6\arcsec$ & $0.728$ & $18.4$ \\
\hline
\end{tabular}
\caption{\label{t_qso} Properties of the background QSO towards which
  the sub-DLA is seen. Columns show the coordinates, emission redshift
  (measured from \MgII\ $\lambda\lambda\ 2296, 2803$ emission in the
  HIRES spectrum) and $R$ band magnitude.}
\end{center}
\end{table}

\begin{table}
\addtolength{\tabcolsep}{-2.pt}
\begin{center}
\begin{tabular}{cccccc}
\hline                                                            
Dataset    & Date obs.   & Exp. Time (s) & Grating & $\lambda_{\rm c}$ (\AA) \\
\hline
LB5H12010  &  6 Nov 2010 & $13905.728$     & G160M & $1589$ \\ 
LB5H13010  & 18 Nov 2010 & $13905.472$     & G160M & $1589$ \\ 
LB5H11010  & 19 Nov 2010 & $13905.376$     & G160M & $1589$ \\ 
LB5H14010  & 24 Nov 2010 & $13905.536$     & G160M & $1623$ \\ 
LB5H15010  & 26 Nov 2010 & $13905.504$     & G160M & $1623$ \\ 
LB5H16010  &  7 Dec 2010 & $13905.472$     & G160M & $1623$ \\ 
\hline
\end{tabular}
\caption{\label{t_COS} Observations of Q~0107$-$0232 with
  HST/COS. Columns show the HST archive dataset name, the date
  observed, total exposure time, grating and central wavelength
  setting used.}
\end{center}
\end{table}

\subsection{COS spectra reduction}

The COS spectra were obtained over a period from the 6th of November
to the 7th of December 2010, as part of the Cycle 17 proposal
11585. They represent a total exposure time of $23$ hours across $30$
orbits.  Two central wavelength settings were taken with the G160M
grating, each using $4$ FP-POS positions to enable complete wavelength
coverage from $1380$ to $1850$~\AA. Details of the exposures are given
in Table~\ref{t_COS}.

We used the CALCOS pipeline\footnote{Version 2.13.6,
  \url{http://www.stsci.edu/hst/cos/pipeline/}} to perform background
subtraction, wavelength calibration and extraction. The default
background extraction smoothing scale of $100$ pixels resulted in poor
background subtraction for our spectra, presumably because the
pipeline was optimised for brighter targets. We found that changing
BWIDTH in the XTRACTAB calibration table from the default value of 100
to $20$ significantly improved the background level such that the flux
in strongly saturated features broader than the COS instrument line
spread profile was consistent with zero.

Wavelength shifts are expected between visits and different wavelength
settings due to temperature differences and uncertainty in the
telescope pointing. The S/N in individual exposures is generally too
low ($\sim 2$ per pixel) to reliably measure the centres of absorption
features. Therefore we combined subsets of exposures grouping by
FP-POS position, by visit (corresponding to a single dataset name in
Table~\ref{t_COS}), by grating central wavelength and by FUV segment
to search for any shifts. Wavelength solutions were consistent across
different visits and FP-POS values, but there are significant
wavelength-dependent shifts between different central wavelength
settings. To correct these, we measured the centroid for common narrow
absorption features where two wavelength settings overlapped, and used
these centres to calculate a wavelength offset as a function of
position. We fitted these offsets with a linear dependence on
wavelength, and then corrected for them such that FUV segment A
$\lambda_c=1627$ matched the FUV segment A $\lambda_c=1589$ setting,
and FUV segment B $\lambda_c=1589$ matched the FUV segment B
$\lambda_c=1627$ setting. The largest shifts applied in this way were
$0.1$ \AA, corresponding to $20$~\kms, but they could result in a
$\sim40$ \kms\ internal shift between the shortest and longest
wavelengths of an exposure. These shifts are given in
Table~\ref{t_wac_shifts}.

The scores of H$_2$ absorption features distributed across the full
spectral range enable a further check of the internal consistency of
the wavelength solution. By measuring the centroid of these features
and comparing to a single-component model of H$_2$ absorption, we
discovered an additional wavelength-dependent shift (shown in
Table~\ref{t_H2_shifts}). The magnitude of this shift is smaller
($\sim 5$ \kms) than that applied above, but still significant when
fitting an absorption system with transitions spread across a large
wavelength range. We removed this shift by subtracting a cubic spline
fitted to the offsets as a function of wavelength position from the
wavelength scale.

To match the zero points of the COS and HIRES wavelength scales, we
compared the \NII~$1084$ and \CI~$945$ features from the $z=0.56$
system in the combined COS spectrum to their expected positions from
the redshifts of the \FeII\ and \MgII\ lines from the same system
measured in the HIRES spectrum. The wavelength zero point of the HIRES
spectrum is known to better than 1 \kms\ relative to narrow Galactic
\CaII\ absorption features seen in the spectrum. Both the \NII\ and
\CI\ appear at redshifts expected from the HIRES \FeII\ and
\MgII\ redshifts, and \NII\ shows a similar component structure
(albeit at the lower COS resolution). We conclude that no correction
to the wavelength zero point of the combined COS spectrum is
necessary.

We also measured the redshift of Galactic absorption features in the
COS spectrum to confirm the zero point of the wavelength solution was
correct. These are all saturated and possibly contain multiple
components, so do not provide a stringent constraint on the zero
point. However, they show no evidence of a systematic offset.

After correcting each exposure for these wavelength shifts we made a
combined spectrum in the following way. First we rebinned each
exposure to a single wavelength scale with pixel width $0.0367$ \AA,
ensuring Nyquist sampling. We used nearest-neighbour binning to
preserve the spectra's noise properties, and checked that this did not
introduce any significant wavelength shifts. The $1\sigma$ uncertainty
on each pixel was estimated empirically as the standard error on the
mean of the contributing pixel fluxes. This is a slight overestimate
of the true uncertainty, as the exposure times were not all
identical. However, the uncertainties measured in this way are
consistent with the standard deviation of the flux in regions free
from absorption, and we believe this is a good estimate of the true
uncertainty.

Since the background level of the COS spectra is low, at small source
count rates the flux distribution may be better described by Poisson
rather than Gaussian statistics. However, in practice we find that for
regions of our spectra with the lowest number of counts -- the cores
of saturated profiles -- uncertainties in the background levels from
the many contributing exposures makes a Gaussian flux distribution a
good approximation.

Finally we estimated the unabsorbed continuum level of the combined
spectrum by fitting spline segments joining regions that appeared free
from absorption. The resulting combined spectrum has a S/N of $10$ per
$\sim 20$~\kms\ resolution element at the continuum and covers a
wavelength range from $1380$ to $1850$~\AA.

\subsection{HIRES spectra reduction}

The HIRES observations were performed on the night of 4th of August
2011. Four 1800\,s exposures were taken using the red cross-disperser
and a 0.861\arcsec\ width slit. Two wavelength settings were used to
cover gaps in the detector.  We used MAKEE to process each exposure,
which subtracts the bias level and the sky background, corrects for
the echelle blaze, generates a wavelength solution by identifying arc
lines to yield a mapping from pixel number to wavelength for each
echelle order, and extracts one-dimensional spectra for each echelle
order. We then used custom-written Python code to coadd the individual
orders for each exposure into a combined spectrum, and to infer the
unabsorbed continuum level by fitting spline segments to regions free
from absorption. The final combined spectrum has a S/N at
$5000$\AA\ of $33$ per $6.67$~\kms\ resolution FWHM, and covers a
wavelength range $3890$ to $8330$~\AA.

\subsection{Imaging}

We acquired $K$ band imaging of a 7\arcmin\,$\times$\,7\arcmin\ field
around Q~0107$-$0232 using the High Acuity Wide field K-band Imager
(HAWK-I) on the Very Large Telescope (VLT) during program
383.A-0402. Five $180\,$s exposures were taken at four offset
positions on the 15th of September 2009. We used the HAWK-I pipeline
recipes to process each exposure to remove the bias level and correct
for sensitivity variations using a flat-field.  An astrometric
solution was measured for each exposure using \textsc{Scamp}
\citep{Bertin06}, then resampled to a common world coordinate system
and coadded all the exposures with \textsc{Swarp} \citep{Bertin02}. We
determined the conversion between the measured counts and the
magnitude by comparison to 2MASS magnitudes for objects in the
field. The limiting magnitude reached is $\sim 23.5$ mag (AB) for a
$3\sigma$ detection of a point source.

\section{Analysis}

\subsection{Line identification}

Most of the transitions associated with the sub-DLA fall inside the
\lya\ forest of the background QSO, and many are blended with
absorption at different redshifts. We identified each absorption
feature in the COS and FOS spectra in the following way. We first
searched for Galactic absorption at the wavelengths of transitions
typically seen in the Galactic interstellar medium (ISM;
\SiII\ $\lambda1526$, \CIV\ $\lambda\lambda1548,\,1550$,
\FeII\ $\lambda1608$, \CI\ $\lambda1657$, and
\MgI\ $\lambda2026$/\ZnII\ $\lambda2026$ were present\footnote{There
  is also absorption at the expected position of \ZnI\ $\lambda2139$,
  redwards of the QSO \lya\ emission. However, since this line is only
  observed in sightlines with \NHI$\gtrsim 10^{21}$~\cmm\ in the Milky
  Way ISM (Daniel Welty, private communication), we identify it as
  \NV\ $\lambda1238$ near the QSO redshift.}). Then we identified
systems by the presence of either \CIV\ ($\lambda\lambda1548,\ 1550$),
\OVI\ ($\lambda\lambda1032,\ 1038$), or \HI\ \lya\ and \lyb, starting
at the emission redshift of the QSO and moving down in redshift to
$z=0$.  Once these systems were identified, we searched for any
further associated metal transitions such as \SiIV, \SiIII, \SiII,
\CIII, \CII.  We found it was necessary to iterate this process
several times, each time including line IDs from previous runs.

The $z=0.56$ sub-DLA was previously identified by \citet{Crighton10}
by its many associated strong metal transitions in the FOS spectrum.
Once we had made plausible identifications for lines at redshifts
other than the $z=0.56$ sub-DLA, we identified metals and molecular
absorption lines from the Lyman and Werner bands for this
system. Finally we assumed any remaining unidentified absorption
features were \lya.  For this paper we focus on absorption features
associated with the $z=0.56$ system. Absorbers at different redshifts
are used only to identify blends with transitions from the sub-DLA.

\subsection{Kinematics and velocity structure of the sub-DLA}

H$_2$ is expected to be found in gas with temperatures less than $\sim
5000$~K -- at higher temperatures molecules are destroyed through
collisional excitation \citep{Shull82}. Therefore we expect the H$_2$
absorption features to be narrow, $<10$~\kms, and the COS spectra will
not resolve the H$_2$-bearing components. H$_2$ components do not
necessarily coincide with the strongest \HI\ or metal line positions
\citep[for example][]{Petitjean02,Noterdaeme10}. However, we use transitions
covered by the higher resolution HIRES spectrum to inform us about the
velocity structure of the absorbing gas, and apply this to H$_2$ and
other transitions only present in the UV spectra.

\begin{figure*}
\begin{center}
\includegraphics[width=0.92\textwidth]{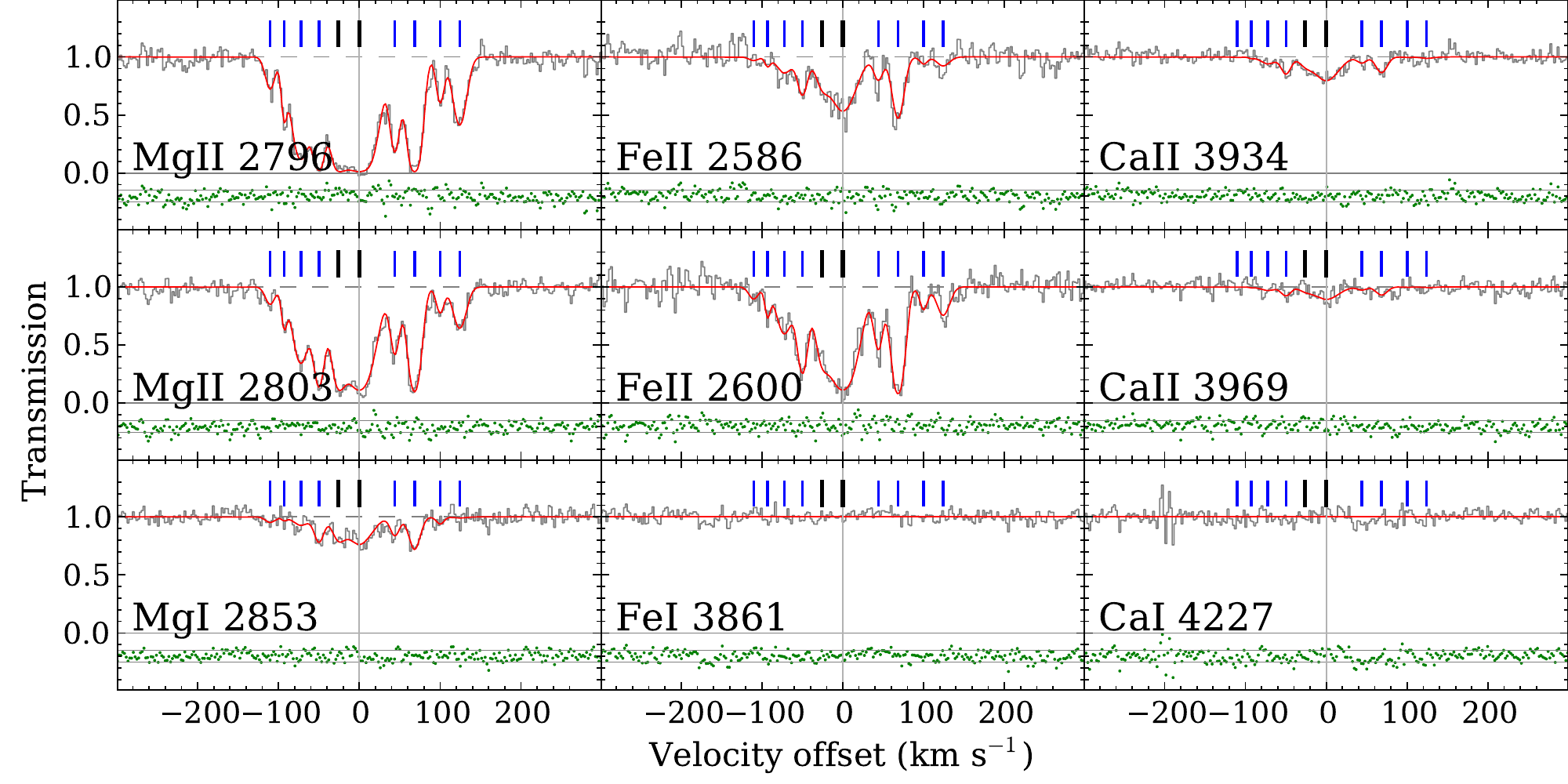}
\caption{\label{f_HIRES} Observed transitions and upper limits at
  $z=0.56$ in the HIRES spectrum. The thin smooth curve shows our
  best-fitting model using a single common velocity structure. The
  darker tick marks show components 5 \& 6 that have associated
  H$_2$. Points distributed around $-0.25$ show the residuals, defined
  as (flux $-$ model) / ($1 \sigma$ error in flux). The two thin lines
  above and below this distribution mark the $1\sigma$ deviation
  levels. Only upper limits are measured for \CaI\ and \FeI.}
\end{center}
\end{figure*}

Figure \ref{f_HIRES} shows the transitions at $z=0.56$ detected in the
HIRES spectrum: \MgII\ ($\lambda\lambda2796,\ 2803$),
\MgI\ ($\lambda2853$), \CaII\ ($\lambda\lambda3934,\ 3969$), and
\FeII\ ($\lambda\lambda2586,\ 2600$). We also measure upper limits on
\AlI, \FeI, \CaI, \NaI, \TiII, and \MnII. We fitted velocity
components and column densities to these transitions using
\textsc{vpfit}\footnote{\url{http://www.ast.cam.ac.uk/~rfc/vpfit.html}}. The
best-fitting values are given in Table~\ref{t_voigt}. A single common
velocity structure spanning $\sim 200$~\kms\ provides a good fit to
all of these transitions, assuming line broadening is dominated by
Gaussian turbulent motions rather than the gas temperature. The best
fitting model is shown in Figure \ref{f_HIRES}.  \CaII\ and \MgI\ have
the lowest ionisation energies ($11.87$ and $7.65$\,eV respectively),
and so {\it a priori} we might expect them to be associated with the
cold environment where H$_2$ is found. However, the photoionisation
analysis in Section~\ref{s_cloudy} indicates that most of the
\MgI\ and much of the \CaII\ probably arises in diffuse, photoionised
gas distinct from the H$_2$.

Component 6 has a Doppler width $b$ ($\equiv \sqrt{2} \sigma$) of
$20$\,\kms, larger than is usually observed in low-ionisation metal
transitions. This, together with the suggestion of correlated
residuals in \MgII\ near the position of this component suggests it is
in fact a blend of two or more narrower components. The quality of
even the HIRES data is not sufficient to constrain the parameters of
such heavily blended components. However, as long as the distribution
of unresolved component widths is not strongly bimodal, the column
density estimates for this component should be accurate
\citep{Jenkins86}. We also measure \NHtwo\ independently of the
velocity model assumed for H$_2$ in Section \ref{s_fhtwo} to ensure
that the velocity model does not strongly bias our measurement of the
molecular fraction.
\begin{table}
\addtolength{\tabcolsep}{-3.8pt}
\begin{center}
\begin{tabular}{clccccccc}
\hline
 \# & Ion  & $\Delta v$ & $\log\,N$& $\sigma_{\log\,N}$ & $b$  & $\sigma_b$ & $z$ & $\sigma_z$  \\
 &  & {\footnotesize (km~s$^{-1}$)} & \multicolumn{2}{c}{\footnotesize ($N$ in cm$^{-2}$)} & \multicolumn{2}{c}{\footnotesize (km~s$^{-1}$)} & & {\footnotesize$\times 10^6$} \\
\hline
1  & \FeII &$-110$& $<12.81$&      & 6.98  & 1.20 & 0.5567157 & 3.7 \\ 
   & \MgI  &      & 10.72   & 0.25 &       &      &           &     \\ 
   & \MgII &      & 12.02   & 0.05 &       &      &           &     \\ 
   & \CaII &      & $<11.48$&      &       &      &           &     \\[\smallskipamount]
2  & \FeII & $-93$& 12.18   & 0.11 & 2.05  & 0.70 & 0.5568053 & 2.0 \\ 
   & \MgI  &      & 10.27   & 0.54 &       &      &           &     \\ 
   & \MgII &      & 12.25   & 0.08 &       &      &           &     \\ 
   & \CaII &      & $<11.40$&      &       &      &           &     \\[\smallskipamount] 
3  & \FeII & $-72$& 12.84   & 0.04 & 11.65 & 0.86 & 0.5569132 & 2.4 \\ 
   & \MgI  &      & 11.08   & 0.14 &       &      &           &     \\ 
   & \MgII &      & 13.02   & 0.02 &       &      &           &     \\ 
   & \CaII &      & 11.32   & 0.12 &       &      &           &     \\[\smallskipamount] 
4  & \FeII & $-50$& 13.09   & 0.03 & 6.94  & 0.54 & 0.5570298 & 1.5 \\ 
   & \MgI  &      & 11.41   & 0.06 &       &      &           &     \\ 
   & \MgII &      & 13.11   & 0.03 &       &      &           &     \\ 
   & \CaII &      & 11.54   & 0.06 &       &      &           &     \\[\smallskipamount] 
5  & \FeII & $-26$& 12.94   & 0.09 & 9.43  & 1.04 & 0.5571530 & 3.5 \\ 
   & \MgI  &      & 11.39   & 0.09 &       &      &           &     \\ 
   & \MgII &      & 13.16   & 0.06 &       &      &           &     \\ 
   & \CaII &      & 11.22   & 0.18 &       &      &           &     \\[\smallskipamount] 
6  & \FeII &  0   & 13.68   & 0.02 & 19.81 & 0.83 & 0.5572885 & 4.6 \\ 
   & \MgI  &      & 11.84   & 0.04 &       &      &           &     \\ 
   & \MgII &      & 13.54   & 0.02 &       &      &           &     \\ 
   & \CaII &      & 12.10   & 0.03 &       &      &           &     \\[\smallskipamount] 
7  & \FeII & +44  & 12.81   & 0.04 &  6.51 & 0.40 & 0.5575174 & 1.2 \\ 
   & \MgI  &      & 11.24   & 0.08 &       &      &           &     \\ 
   & \MgII &      & 12.72   & 0.02 &       &      &           &     \\ 
   & \CaII &      & 11.05   & 0.17 &       &      &           &     \\[\smallskipamount] 
8  & \FeII & +68  & 13.42   & 0.02 &  8.41 & 0.25 & 0.5576435 & 0.8 \\ 
   & \MgI  &      & 11.60   & 0.04 &       &      &           &     \\ 
   & \MgII &      & 13.25   & 0.02 &       &      &           &     \\ 
   & \CaII &      & 11.57   & 0.06 &       &      &           &     \\[\smallskipamount] 
9  & \FeII & +100 & 12.20   & 0.10 &  5.08 & 0.76 & 0.5578081 & 2.3 \\ 
   & \MgI  &      & 10.74   & 0.25 &       &      &           &     \\ 
   & \MgII &      & 12.13   & 0.02 &       &      &           &     \\ 
   & \CaII &      & $<11.44$&      &       &      &           &     \\[\smallskipamount] 
10 & \FeII &+124  & 12.51   & 0.08 &  9.74 & 0.56 & 0.5579336 & 1.8 \\ 
   & \MgI  &      & 10.15   & 1.03 &       &      &           &     \\ 
   & \MgII &      & 12.57   & 0.02 &       &      &           &     \\ 
   & \CaII &      & $<11.56$&      &       &      &           &     \\ 
\hline
\end{tabular} 
\caption{\label{t_voigt} Best-fitting Voigt profile parameters for
  each component in transitions for the $z=0.56$ system that are
  covered by the HIRES spectrum.}
\end{center}
\end{table}

\subsection{UV transitions for the sub-DLA}
\begin{figure}
\begin{center}
\includegraphics[width=0.47\textwidth]{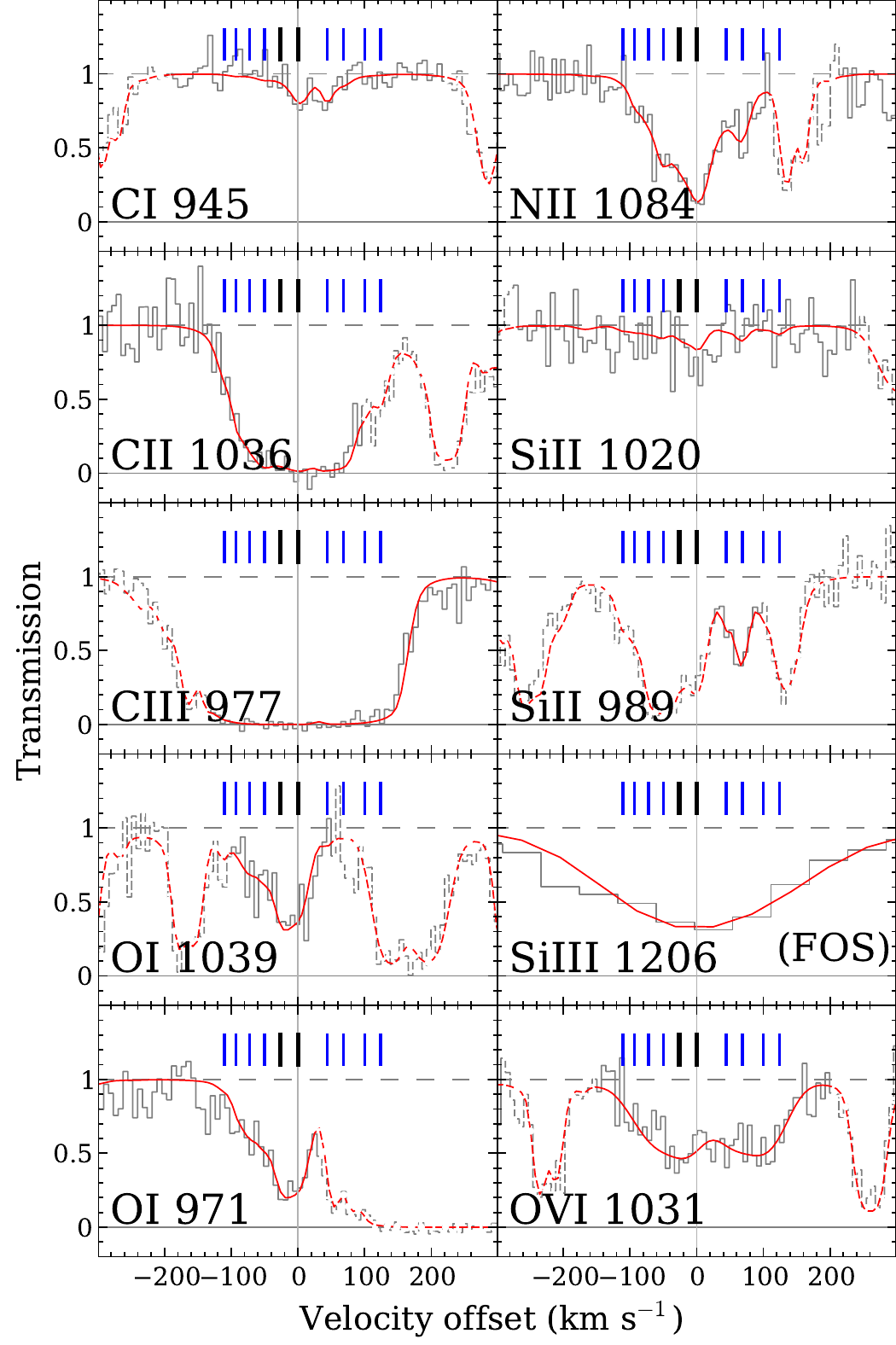}
\caption{\label{f_COS} Transitions for the $z=0.56$ sub-DLA in the COS
  and FOS UV spectra. The smooth red line shows Voigt profile models
  of the data. Absorption that is not due to the named transition in
  each panel (usually from H$_2$) is shown by dashed lines. Tick marks
  show the component positions from Figure~\ref{f_HIRES}, dark ticks
  show components that have associated H$_2$ absorption. All
  transitions are covered by the COS spectra, apart from \SiIII, which
  is covered by the lower resolution FOS spectrum.}
\end{center}
\end{figure}

We apply the \MgII\ velocity structure to models fitted to transitions
observed in the lower resolution COS and FOS spectra. Using this
velocity structure we were able to match the \NII, \SiII\ and
\OI\ profiles by varying the component line widths and column
densities.  Several of the COS transitions that have measurable
absorption and are not saturated or heavily blended with unrelated
systems are shown in Figure~\ref{f_COS}. When fitting the COS spectra
we use the tabulated line spread function provided by
STScI\footnote{\url{http://www.stsci.edu/hst/cos/performance/spectral_resolution/}},
linearly interpolated to the wavelength at the centre of each fitting
region.  We also measured column densities using the apparent optical
depth (AOD) method (which assumes the transition is optically thin,
\citealt{Savage91}), including a $5$ per cent uncertainty in the
continuum level.  As the individual components are not resolved by the
COS spectra, we quote these AOD measurements and give the total column
densities for all components in aggregate. For transitions \CI, \NII,
\OI, \OVI\ we were able to directly compare column densities measured
using both Voigt profile fitting and the AOD method. In each case they
are consistent with one another. \CII, \CIII\ and \SiIII\ are
saturated, and lower limits are measured using the AOD method. The FOS
spectrum provides an upper limit on $N_\textrm{\SiIV}$. Table~\ref{t_logN_UV} gives measurements and uncertainties,
lower and upper limits for all of the transitions in the UV spectra.

\begin{figure}
\begin{center}
\includegraphics[width=0.48\textwidth]{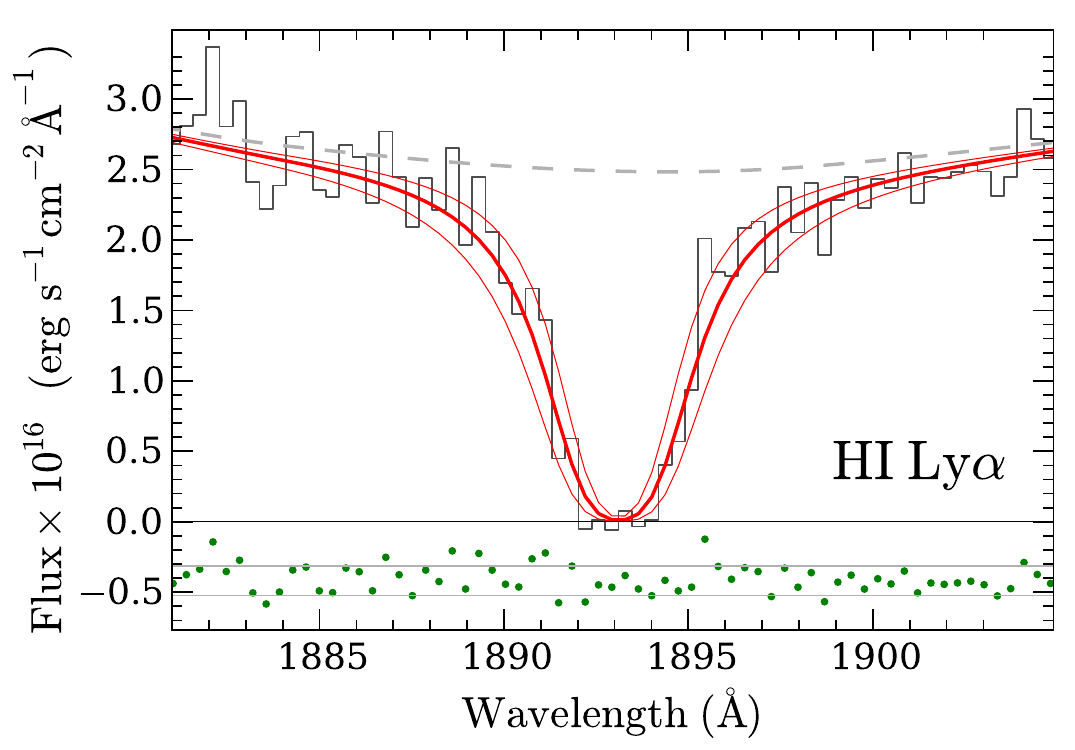}
\caption{\label{f_NHI} Constraints on \NHI\ from the
  \lya\ transition. The histogram shows the data, the dashed line the
  continuum, and the lower green points the residuals as defined in
  the caption of Figure~\ref{f_HIRES}. The thick red solid curve shows
  the best-fitting \NHI\,$=10^{19.5}$\,\cmm\ for a single component
  with $b=20$\,\kms, and the thinner upper and lower solid curves show
  \NHI\,$=10^{19.3}$ and $10^{19.7}$\,\cmm.}
\end{center}
\end{figure}

The damping wings measured at \lya\ in the FOS spectrum constrain
\NHI\ $=10^{19.5 \pm 0.2}$~\cmm, where the error is dominated by the
systematic uncertainty in the continuum level (see
Figure~\ref{f_NHI}).
\begin{table}
\renewcommand{\arraystretch}{1.2}
\begin{center}
\begin{tabular}{ccc}
\hline
 Ion &  Transition $\lambda$ (\AA) & $\log_{10} N$ (cm$^{-2}$) \\
\hline
   \HI    & 1215 &  $19.50^{+0.20}_{-0.20}$  \\ 
   \CI    &  945 &  $13.53^{+0.24}_{-0.77}$  \\ 
   \CII   & 1036 &  $> 14.8$              \\
   \CIII  &  977 &  $> 14.3$              \\
   \NI    & 1135 &  $< 14.4$              \\ 
   \NII   & 1084 &  $14.73^{+0.17}_{-0.19}$ $^{\mathrm{a}}$   \\
   \OI    & 1039 &  $15.53^{+0.24}_{-0.25}$  \\
   \OVI   & 1031 &  $14.60^{+0.16}_{-0.24}$  \\ 
   \SiII  & 1020 &  $14.79^{+0.23}_{-0.64}$  \\ 
   \SiIII & 1206 &  $> 13.7$              \\ 
   \SiIV  & 1393 &  $< 13.1$              \\ 
\hline
\end{tabular}
\caption{\label{t_logN_UV} Total column densities for transitions in
  the $z=0.56$ system observed in the COS and FOS spectra. \NHI\ is
  calculated from the damping wings at \lya\ in the FOS
  spectrum. \CII, \CIII\ and \SiIII\ are saturated, and lower limits
  are calculated using the AOD method. The \NI\ and \SiIV\ values are
  $5\sigma$ upper limits. The remaining values were calculated using
  the apparent optical depth of the transition with rest wavelength in
  the second column. Uncertainties given for these values are $1
  \sigma$ and include a $5$ per cent uncertainty in the continuum
  level, which generally dominates the statistical
  uncertainty. ${^\mathrm{a}}$: velocity models for \NII\ with a
  saturated central component allow higher column densities than this
  value and are still compatible with the data, but only by using a
  more complicated velocity structure than that fitted to the HIRES
  transitions.}
\end{center}
\end{table}

\subsection{H$_2$ velocity structure}
We measure H$_2$ transitions from the $J=0-3$ rotational levels, with
upper limits on $J=4$ and $5$. The asymmetric profiles for many of the
H$_2$ lines suggest there is more than one absorbing component. We
were also unable to successfully fit the equivalent widths of the
transitions using a curve of growth analysis with a single
component. Therefore we fitted two H$_2$ components, with redshifts
close to those of the two central strong metal components at
$-26$~\kms\ and $0$~\kms\ (components 5 and 6 in
Table~\ref{t_voigt}). These are clearly separated in the resolution
$\sim 6$~\kms\ HIRES spectra, but blended at the instrumental line
profile of COS.  However, the large number of transitions over a range
of oscillator strengths allow us to constrain velocity structure below
the instrumental resolution. As it is not uncommon for H$_2$ to be
significantly offset from the strongest metal absorption -- indeed, in
Section~\ref{s_threephase} we show that the H$_2$ is probably produced
in a different environment to most of the metal lines -- we allow the
redshifts of each H$_2$ component to vary in our fitting procedure.

We experimented with fitting the two components using \textsc{vpfit},
and found there were large degeneracies between the Doppler $b$
parameter and column density. One way to robustly explore the
$b$-$N$-$z$ parameter space for the two components is to generate
large grids of likelihood values as a function of the fitted
parameters over plausible regions of parameter space. However, in our
case this proved to be prohibitively expensive
computationally. Instead we used a Monte-Carlo Markov Chain (MCMC)
technique to sample parameter space. This samples parameter values in
proportion to the likelihood value at any point in parameter
space. Thus from a set of initial parameter positions, a `chain' of
parameter values is generated by a stochastic walk through parameter
space with distributions approximating the Bayesian posterior
probability for each parameter.

\begin{figure*}
\begin{center}
\includegraphics[width=0.92\textwidth]{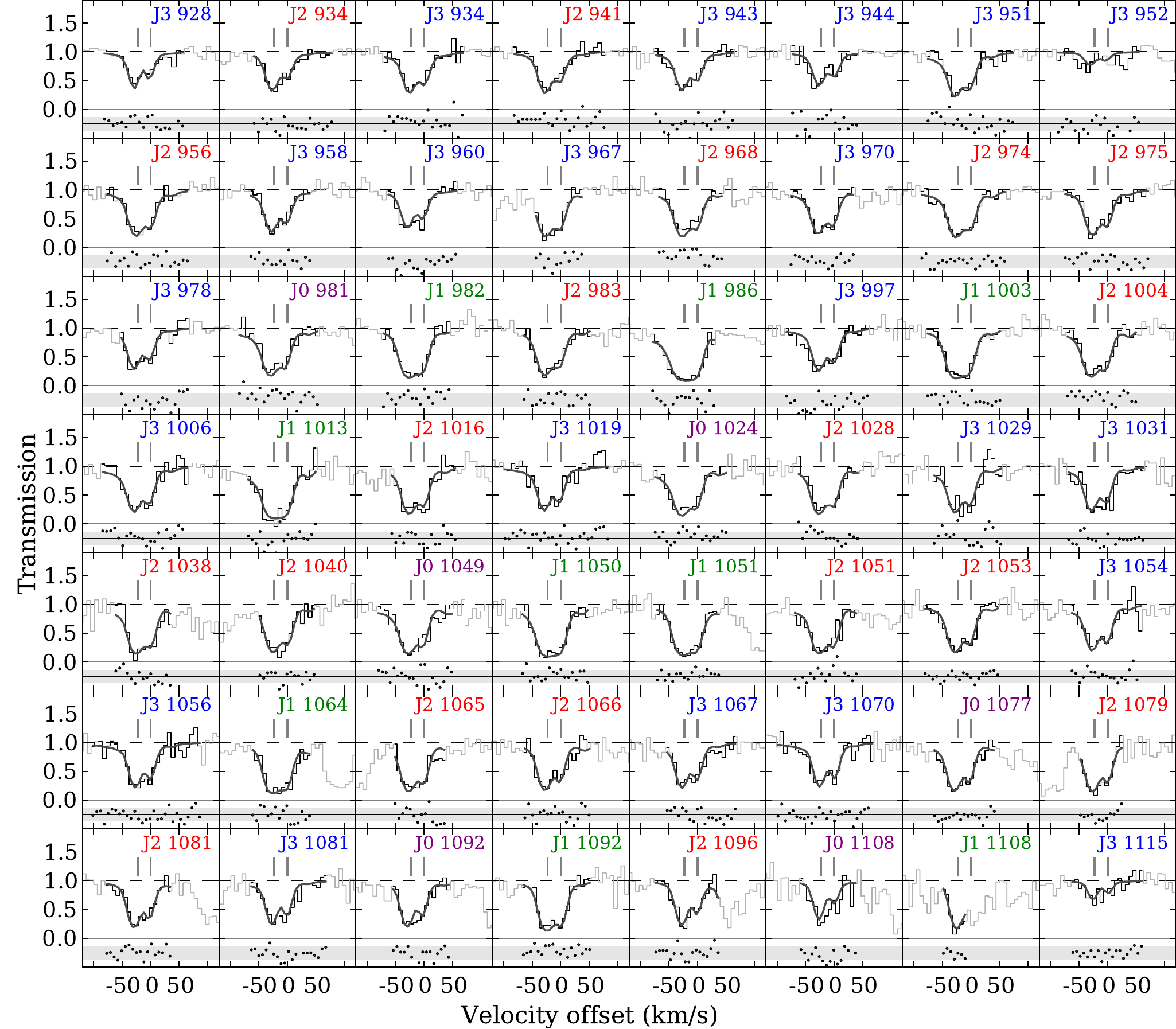}
\caption{\label{f_Htwo_model} The H$_2$ absorption model generated
  using parameters that maximise the likelihood of the data. The
  histogram shows the flux divided by the continuum (the darker
  portions were used to calculate the likelihood), the smooth curves
  show the model. The residuals -- (data - model) / $1\sigma$ -- are
  shown centred on $-0.25$, scaled such that the light gray lines
  above and below $-0.25$ are the $\pm 1 \sigma$ range. A two
  component model with a single $b$-parameter for each component and
  redshifts corresponding to components 5 \& 6 in Table~\ref{t_voigt}
  reproduces the data well.}
\end{center}
\end{figure*}

We generated posterior parameter distributions using the package
\textsc{Emcee} \citep[The MCMC Hammer,][]{ForemanMackey12}. We fitted
for each component's redshift and $b$ parameter, and for the column
density of each rotational level using the 56 transitions shown in
Figure~\ref{f_Htwo_model}. All transitions for a component were
constrained to have the same $b$ value, which is generally observed to
be the case in local sites of H$_2$ absorption for at least $J$ levels
$<3$ \citep{Spitzer74}. Fitting transitions from each rotational level
individually also gives $b$ parameters consistent with a single
value. Even after correcting the wavelength scale, residual wavelength
shifts of $\sim 1$~\kms\ remain, so we also allowed a small wavelength
shift for each fitted region. Thus we fit for 8 column densities, two
redshifts, two $b$ parameters, and one wavelength offset for each of
the 56 regions resulting in a total of 68 parameters. 

Table~\ref{t_logN_Htwo} gives the parameter estimates and $1\sigma$
errors for the velocity offsets (with respect to metal components 5
and 6), $b$ parameters, and H$_2$ column densities for each rotational
level. The $1\sigma$ regions are determined by marginalising over all
other parameters and finding the narrowest region that encompasses
$68.3$ per cent of the samples.  We choose the parameter estimates to
be at the centre of these $1\sigma$ regions. The absorption model with
the set of parameters that maximises the likelihood is shown in Figure
\ref{f_Htwo_model}.
\begin{table}
\addtolength{\tabcolsep}{-2.5pt}
\begin{center}
\begin{tabular}{cccc}
\hline
$J$ & $\log_{10} N$ (cm$^{-2}$) & $b$ (\kms) & $\delta v$ (\kms) \\
\hline
& \multicolumn{3}{c}{Component 5}  \\
0 & $16.17\pm 0.25$ & $ 6.7\pm 0.6$ & $3.5\pm 0.6$  \\
1 & $17.05\pm 0.28$ &               &               \\
2 & $16.19\pm 0.19$ &               &               \\
3 & $15.77\pm 0.12$ &               &               \\
4 & $<14.5$         &               &               \\
5 & $<14.5$         &               &               \\
& \multicolumn{3}{c}{Component 6} \\
0 & $15.63\pm 0.39$ & $4.3\pm 0.7$  & $4.3\pm 0.7$  \\ 
1 & $16.42\pm 0.40$ &               &               \\ 
2 & $15.65\pm 0.25$ &               &               \\ 
3 & $15.47\pm 0.18$ &               &               \\ 
4 & $<14.5$         &               &               \\ 
5 & $<14.3$         &               &               \\ 
\hline
\end{tabular}
\caption{\label{t_logN_Htwo} Column densities, $b$ values and velocity
  offsets for the two components showing H$_2$ based on Monte Carlo
  Markov Chain fitting.  Errors are $1 \sigma$ and the $J=5$ \& 6
  values are $5\sigma$ upper limits. The velocity offsets are from the
  redshifts of metal components 5 and 6, given in Table~\ref{t_voigt}.}
\end{center}
\end{table}

\section{Absorption system properties}
\subsection{Metallicity}
The metallicity, $Z$, can be estimated from an element X using the log
of the ratio of the abundance of element X in the absorber,
$N_\textrm{X}/N_\textrm{H}$, to the solar abundance
\begin{equation}
[\mathrm{X}/\mathrm{H}] \equiv \log_{10}
\frac{(N_\textrm{X}/N_\textrm{H})_\mathrm{obs}}{(N_\textrm{X}/N_\textrm{H})_{\sun}}.
\end{equation}
Due to a charge transfer between O and H we expect the ratio of the
number densities $n_{\rm OI}/n_{\rm O}$ and $n_{\rm HI}/n_{\rm H}$ to
be the same \citep{Field71}, provided the majority of O is in the form
of \OI\ and \OII. In the presence of many high energy ionising
photons, this is no longer true due to different absorbing cross
sections of \OI\ and \HI\ \citep[see][]{Prochter10}, but the absence
of \SiIV\ argues against a hard radiation field for this system, and
the best fitting \textsc{Cloudy} models do not predict significant
amounts of O in higher ionisation states (\OVI\ is seen, but we argue
this occurs in a hotter, collisionally-ionised phase). Oxygen also
shows little depletion ($< 0.3$~dex) onto dust grains across a range
of environments in the ISM of the Milky Way \citep{Jenkins09}, so
should provide a good estimate of the metallicity.

We find [\OI/\HI]\,$=-0.72\pm 0.32$, or $\sim 0.19\,Z_\odot$.  For the
photoionisation analysis in Section~\ref{s_cloudy} we assume the
metallicity is the same across the entire complex. In one of the few
cases where the metallicity has been measured for individual
components in a single absorption system, metallicity differences of a
factor of ten have been observed \citep{Prochter10}. However, given
that the dispersion in $N_\textrm{\MgII} / N_\textrm{\FeII}$ across the
system is not excessively large (the largest log difference between
components is $0.38$), the assumption of a constant metallicity seems
reasonable. In particular, the two components with H$_2$ do not show
significantly different ion abundance ratios compared to the entire
system.

\subsection{Dust}
\label{s_kappa}
\citet{Noterdaeme08} have found a correlation between the presence of
dust and the likelihood of observing H$_2$ in DLAs, consistent with
the main formation mechanism for H$_2$ being on the surface of dust
grains. We can measure the dust content by comparing elements known to
deplete strongly onto dust grains (Fe, Mg) to those with low depletion
(O). We assume negligible ionisation corrections, but applying
corrections from the best-fitting \textsc{Cloudy} model in
Section~\ref{s_cloudy} does not change our conclusions. We find
[Fe/O]~$= -0.25^{+0.21}_{-0.29}$ and [Mg/O]~$= -0.38 \pm 0.28$ for the entire
system, indicating mild dust depletion. If we also assume solar
abundance ratios we can estimate the dust-to-gas ratio normalised by
the value in the solar neighbourhood as
\begin{equation}
  \kappa = 10^\mathrm{[X/H]} ( 1 - 10^\mathrm{[Fe/X]} )\, ,
\end{equation}
where X is an element that does not deplete strongly onto dust
\citep[for a derivation of this expression see the appendix
  of][]{Wolfe03}. Using oxygen gives $\log_{10} \kappa < -0.44$,
compatible with values found in other higher redshift systems showing
H$_2$ \citep{Ledoux03}.

\subsection{Temperature constraints from line widths}

The linewidths of absorption components in the HIRES spectra can be
used to constrain the temperature of the gas using the relation $b =
\sqrt{2kT/m}$, where $m$ is the mass of the ion and $T$ is the
temperature.  This constraint is an upper limit, as there can be
large-scale turbulent motions in addition to thermal broadening, or
the line may not be resolved. Indeed, we fitted each component in
\MgII, \FeII, and \CaII\ with a single $b$ parameter value across all
three transitions, consistent with turbulent broadening dominating
over thermal broadening. Due to the relatively large masses for these
elements, only component 2 gives a constraining upper limit of
$6\,000$~K, typical of temperatures in the warm neutral medium in the
Milky Way ISM.  The H$_2$ linewidths give upper limits to the
temperature of $6\,500$~K and $5\,400$~K for the blue and redder
components, but we argue below that the physical conditions in the
H$_2$ gas are probably different to those of the gas where most of the
metal lines arise. It is also possible that the H$_2$ widths are
substantially broadened due to turbulent motions.

In conclusion, there are no strong temperature constraints from the
linewidths. The relative column densities of the H$_2$ rotational
levels provide an independent measure of the temperature, discussed in
the next section.

\subsection{H$_2$ excitation temperature}
\label{s_Tex}
The ratios of H$_2$ column densities in different rotational levels
can be expressed as excitation temperatures, assuming a Boltzmann
distribution across the levels \citep[see][]{Draine11}:
\begin{equation}
  \frac{N_J}{N_{J=0}}=\frac{g_J}{g_{J=0}} \exp\left(\frac{-B_vJ(J+1)}{T_{J0}}\right).
\end{equation}
Here $N_J$ is the column density for molecules in rotational state
$J$, and $g_J \equiv (2J + 1)(2I + 1)$, where $I=0$ if $J$ is odd or 1
if $J$ is even, is the statistical weight of $J$. $B_v = 85.36$~K and
$T_{J0}$ is the excitation temperature from level $J$ to $J=0$.

\begin{figure}
\begin{center}
\includegraphics[width=0.485\textwidth]{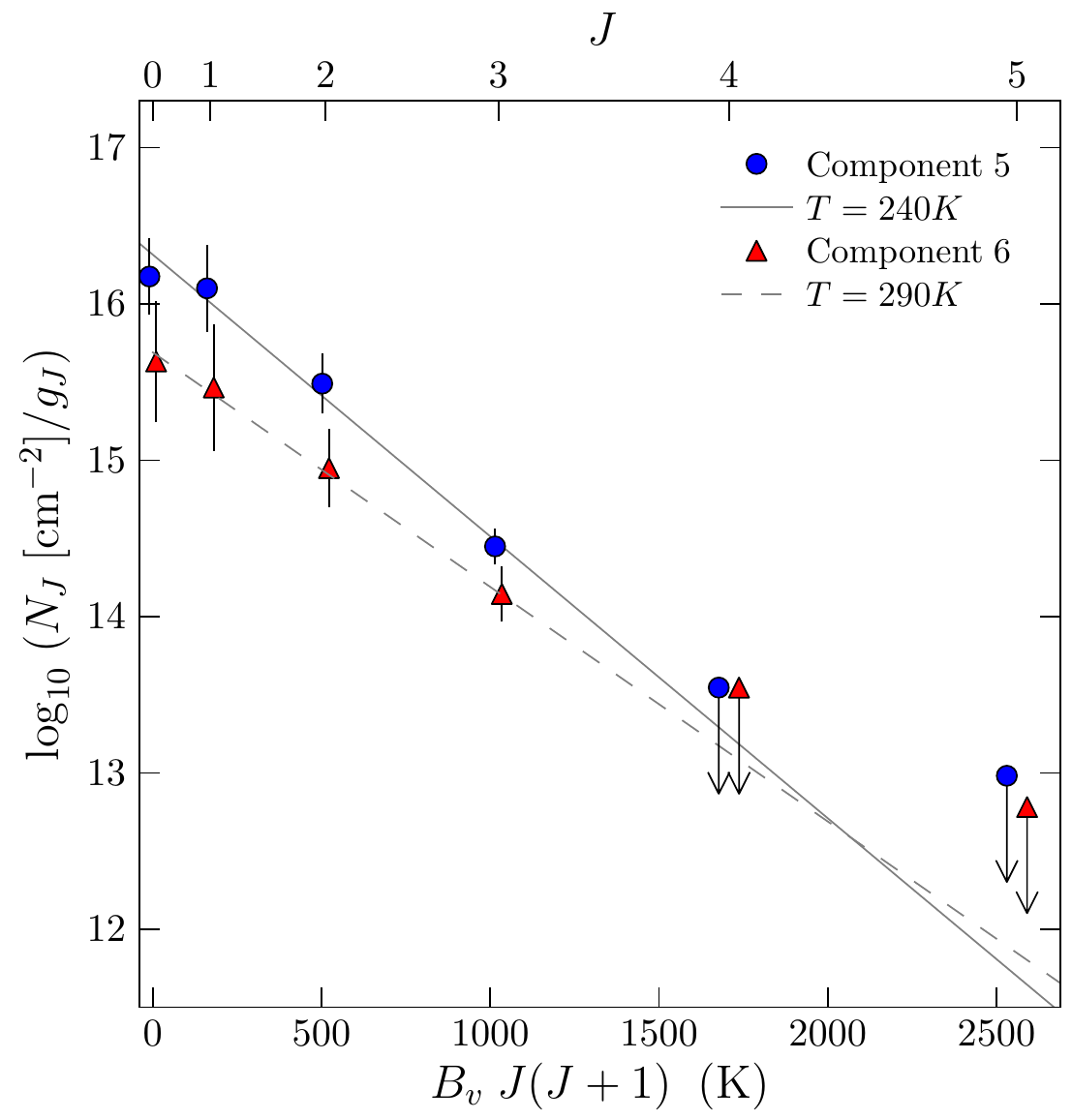}
\caption{\label{f_Tex} Excitation diagram for the two H$_2$
  components. The error bars show the $1 \sigma$ uncertainties. The
  slope of a line joining each pair of points is inversely
  proportional to the excitation temperature between those two $J$
  levels. An illustrative temperature is shown for each component, but
  the population levels are not expected to follow a Boltzmann
  distribution characterised by a single temperature.}
\end{center}
\end{figure}

Fig.~\ref{f_Tex} shows an excitation diagram for the column densities
of the $J=0 - 3$ transitions for the two H$_2$ components. If the
collisional timescale for the $J=0$ and $J=1$ transitions is much
shorter than the photodissociation timescale, which occurs above
densities of $\sim 100$~\cmmm\ when H$_2$ is sufficiently
self-shielded from dissociating photons, then $T_{10}$ represents the
kinetic temperature of the gas \citep[see for
  example][]{Dalgarno73}. The $z=0.56$ system is likely only partially
self-shielded, but assuming it satisfies these requirements we find a
lower limit on $T_{10}$ for each component at $1\sigma$ ($2\sigma$)
limits of $123$~K ($64$~K) for component 5 and $77$~K ($37$~K) for
component 6. Two illustrative temperatures corresponding to the
populations for $J= 0 - 3$ in each component are shown in
Fig.~\ref{f_Tex}. However, different physical processes affect the
populations of these levels \citep{Jura75b}, so it is not expected
that a single temperature should match all four levels.

\subsection{\textsc{Cloudy} modelling}
\label{s_cloudy}
In this section we attempt to generate a simple single-cloud model
illuminated by a UV radiation field that can reproduce all the
observed column densities. We compare to the total column densities
for all components, since the individual component columns are not
well constrained for the \OI, \CI, and \SiII\ transitions or the
saturated transitions (\CII, \CIII\ and \SiIII). Given the large range
of transitions present with widely differing ionisation energies, it
is likely that there are several different phases present, and a
single cloud model is unlikely to be able to reproduce all the
observed species. Below we find that a single model can reproduce the
majority of the low-ionisation metal transitions, but
Section~\ref{s_threephase} shows that multiple phases with different
densities and temperatures are required to explain all the absorption.

We use models generated with version 8.01 of \textsc{Cloudy}, last
described by \citet{Ferland98}, to estimate the physical conditions in
the absorption system. All models assume solar abundance ratios,
constant gas density, and an absorbing geometry of a thin slab
illuminated on one side by an incident radiation field perpendicular
to the slab surface.  The radiation field includes the cosmic
microwave background at the redshift of the absorber. We then compare
four scenarios: a cloud in an intergalactic medium-like (IGM)
environment, in an ISM-like environment, close to a starburst galaxy,
and illuminated by an AGN-dominated spectrum. We chose the
AGN-dominated spectrum to estimate the effect of a nearby AGN that may
be present in one of the galaxy candidates described in
Section~\ref{s_gal}, and to see if a spectrum with more high-energy UV
photons can produce the observed \OVI\ column density in addition to
that of the low ionisation transitions. The IGM-like model is free of
dust with a radiation field given by the UV background spectrum from
\citet{Haardt12}, including contributions from quasars and
star-forming galaxies at the redshift of the absorber. It has a
radiation field strength at $912$\,\AA,
$J_{\nu,\,\mathrm{IGM}}^{\,912} =
6.08\times10^{-23}$erg~s$^{-1}$cm$^{-2}$Hz$^{-1}$sr$^{-1}$. The
ISM-like models have a radiation field similar to the Galactic ISM
($J_\nu^{\,912} \sim 400 J_{\nu,\,\mathrm{IGM}}^{\,912}$), which is
dominated at UV wavelengths by the spectral shape of hot stars, and a
dust grain composition similar to that measured in the ISM. Even
though the ISM-like models use solar relative abundances, the gas
phase abundance ratios are substantially different from solar due to
differential depletion of metals onto grains. The starburst
\textsc{Cloudy} models assume the absorbing cloud is 10\,kpc from the
galaxy and an escape fraction for UV light of 3 per cent, in addition
to the IGM-like radiation field described above, without dust grains.
They have $J_\nu^{\,912}\sim 2700 J_{\nu,\,\mathrm{IGM}}^{\,912}$.
The starburst galaxy spectrum used was generated using
\textsc{Starburst99} \citep{Leitherer99} for a star formation rate of
$20$\,\msunyr. The AGN models use the default tabulated AGN spectrum
from \textsc{Cloudy} with a normalisation $J_\nu^{\,912} \sim 3000
J_{\nu,\,\mathrm{IGM}}^{\,912}$ and do not include dust grains.

For each scenario we generate a grid of models for a range of
ionisation parameters, metallicities and total \NHI. We estimate the
ionisation parameter $U$, defined as the ratio of the densities of
ionising photons to hydrogen atoms, using the observed total column
density ratios $N_\textrm{\MgI}/N_\textrm{\MgII}$,
$N_\textrm{\NI}/N_\textrm{\NII}$, and
$N_\textrm{\SiII}/N_\textrm{\SiIII}$. Using ratios of ionisation states
for the same element avoids any effects that might alter the column
densities of ions for different elements in different ways, such as
non-solar abundance ratios or differential dust depletion.  We
generate the likelihood of each parameter ($U$, $Z$ and $\NHI$) for
the grid of models based on the observed ratios, and include a
Gaussian prior on the metallicity centred on the [\OI/\HI] metallicity
measurement with width $\sigma$ equal to the $1 \sigma$ uncertainty on
the metallicity.  For all three scenarios, only a relatively narrow
range of $U$ values correctly reproduces the observed ratios. The
likelihoods are only weakly dependent on the total \NHI; we assume
\NHI\,$=10^{19.5}$\,\cmm, which results in models that best reproduce
the observed metal column densities.

Once we have found the most likely $U$ value, we compare the predicted
column densities to the observed transitions with measurements or
limits, and assess which scenario reproduces the observations best. We
first compare to the AGN models. These are the only models with a hard
enough spectrum to produce sufficient $N_\textrm{\OVI}$ to match the
observed value. However, at the same time they overpredict the amount
of \SiIV, \FeII, \MgI\ and \MgII\ by one or more orders of
magnitude. Thus it is more likely the \OVI\ arises in a
collisionally-ionised phase separate from the low-ionisation
transitions, as is observed in other systems \citep[for
  example][]{Fox07,Ribaudo11}, and we do not compare to the high-ionisation
species (\SiIV, \OVI) for the remaining models.  

\begin{figure}
\begin{center}
\includegraphics[width=0.4\textwidth]{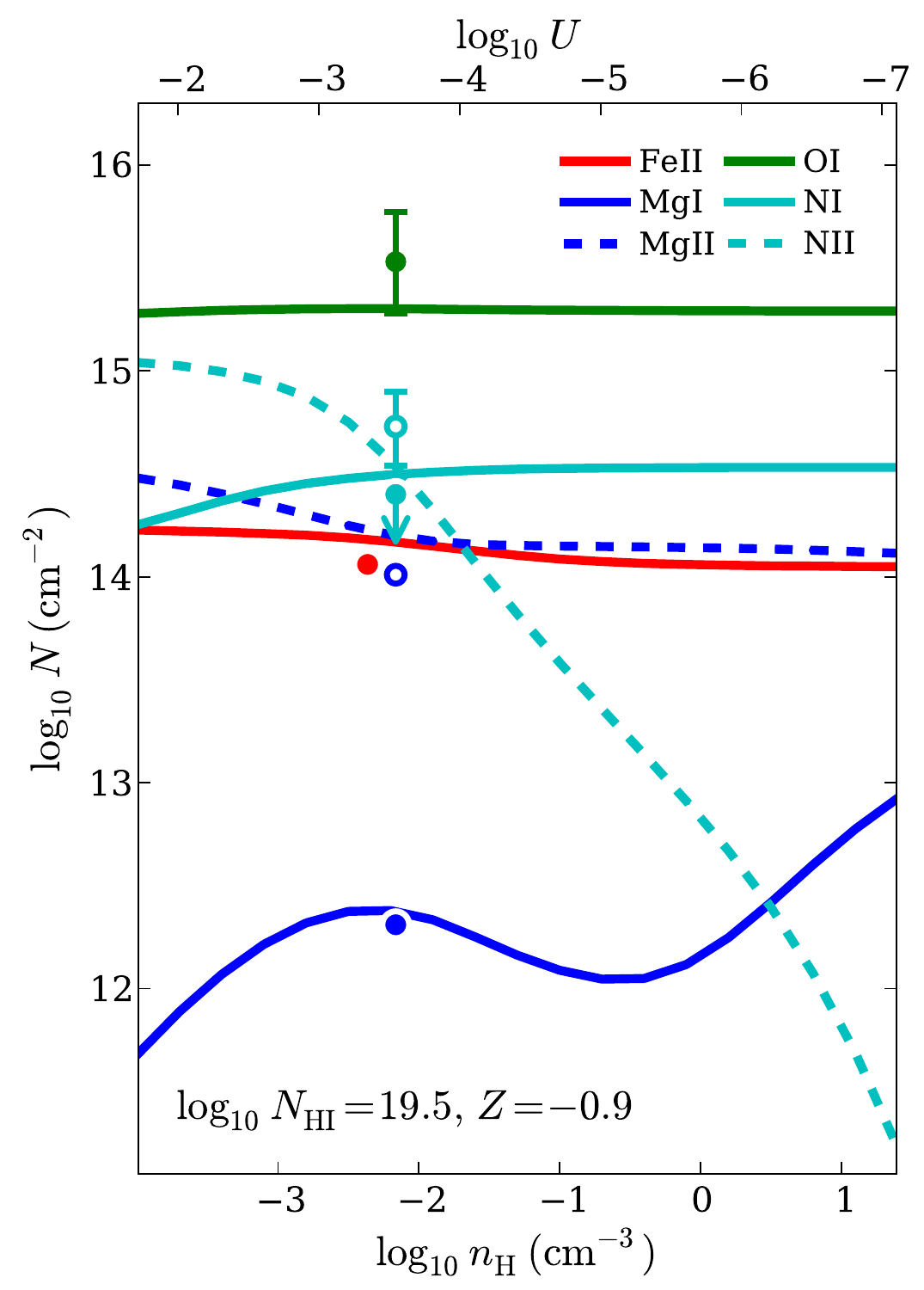}
\caption{\label{f_cloudy1} \textsc{Cloudy} predictions (lines) and
  observed column densities (points) for \MgI, \MgII, \NI, \NII,
  \OI\ and \FeII\ for the IGM-like model with $Z =
  0.13\,Z_\odot$. Points are plotted at the most likely $U$ values
  based on constraints from the $N_\textrm{\MgI}/N_\textrm{\MgII}$,
  $N_\textrm{\NI}/N_\textrm{\NII}$, and
  $N_\textrm{\SiII}/N_\textrm{\SiIII}$ ratios (\FeII\ is offset for
  clarity). Filled points correspond to solid lines of the same
  colour, open points to dashed lines. Errors on the column densities
  are smaller than the marker sizes. Given the uncertainties in the
  shape of the ionising spectrum and uncertain relative abundances,
  the predicted column densities are very close to the observed
  values.}
\end{center}
\end{figure}

\begin{figure}
\begin{center}
\includegraphics[width=0.4\textwidth]{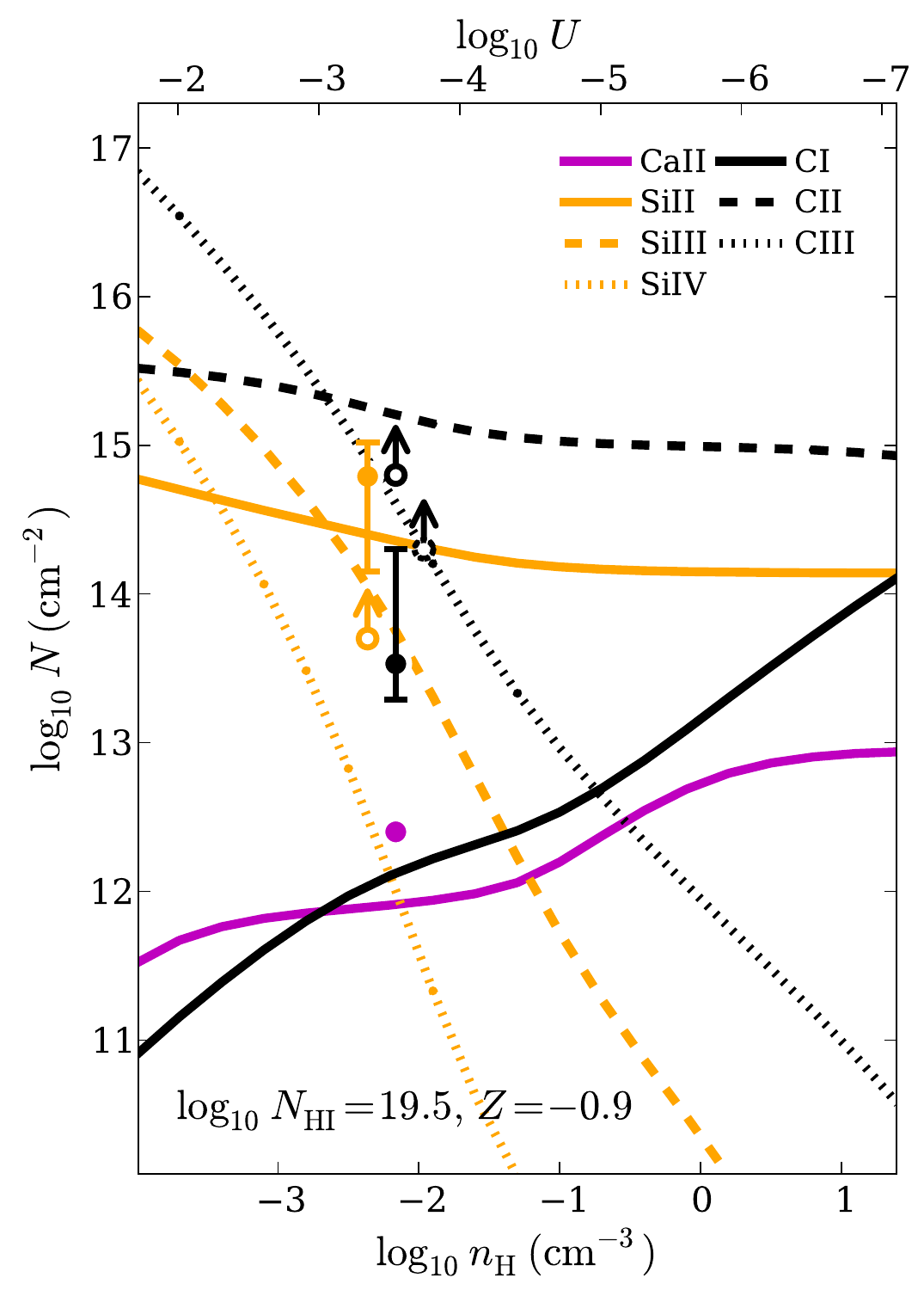}
\caption{\label{f_cloudy2} \textsc{Cloudy} predictions (lines) and
  observed column densities (points) for \CI, \CII, \CIII, \SiII,
  \SiIII\ and \CaII\ for the IGM-like model with $Z =
  0.13\,Z_\odot$. Points are plotted at the most likely $U$ values
  based on constraints from the $N_\textrm{\MgI}/N_\textrm{\MgII}$,
  $N_\textrm{\NI}/N_\textrm{\NII}$, and
  $N_\textrm{\SiII}/N_\textrm{\SiIII}$ ratios (Si and \CIII\ points are
  offset for clarity). Filled points correspond to solid lines of the
  same colour, open points to dashed lines, and open points with
  dotted edges to dotted lines. Where error bars are not shown they
  are smaller than the marker size. With the exception of \CI\ and
  \CaII, the column densities are reproduced well. As discussed in
  Section~\ref{s_threephase}, the excess $N_{\CI}$ is likely due to
  \CI\ being in a self-shielded cold phase where the H$_2$ and
  possibly part of the \CaII\ resides.}
\end{center}
\end{figure}

From the mild depletions measured in Section \ref{s_kappa}, we already
expect that the ISM-like case will not match the observed
abundances. The most likely model does indeed underpredict the
\FeII\ abundance by more than an order of magnitude, and \CaII\ by
many orders of magnitude, as both are expected to be heavily depleted
onto dust grains. This confirms that the depletion pattern in the
$z=0.56$ sub-DLA is different from that in the Milky Way ISM. This
model also underpredicts \CI. The starburst scenario fails to
reproduce the observed $N_\textrm{\NI}/N_\textrm{\NII}$, and also
severely underpredicts \CaII, \FeII\ and \CI. The IGM-like model gives
the best fit to the observed data, and its predictions along with
observed column densities are shown in Figures~\ref{f_cloudy1} \&
\ref{f_cloudy2}. Figure~\ref{f_cloudy1} shows that column densities
for \OI, \MgI, \MgII, \FeII, \NI\ and \NII\ are reasonably well
matched.  The remaining small deviations from the predictions could be
due to a slightly different incident UV continuum from the one
assumed, or enhanced or depleted elemental abundances relative to the
solar values assumed. For example, $0.2$ dex less $N_\textrm{\NII}$ is
observed than is predicted. This may be due to a nitrogen
underabundance, often observed in similar \NHI\ systems at low
redshifts \citep[for example][]{Battisti12} and in DLAs at higher
redshifts \citep{Pettini02, Prochaska02}. Figure~\ref{f_cloudy2} shows
that \SiII, \SiIII, \CII\ and \CIII\ columns are reproduced
well. However, there is still $0.5$ dex too little $N_\textrm{\CaII}$
and 1 dex too little $N_\textrm{\CI}$ predicted. In all three
scenarios, we also find that the \NHtwo\ predicted is more than an
order of magnitude below the observed value. In
Section~\ref{s_threephase} we suggest a scenario to explain this
discrepancy between the models and observations.

We also ran \textsc{Cloudy} models using constant pressure clouds instead of
constant density. The motivation for these was to simultaneously
include cool, lower density gas at the edge of the cloud, and higher
density, cold $\sim100$\,K gas at the core of the cloud where H$_2$
can survive. In these models we also included contributions from
cosmic rays, which can be important for cold molecular
regions. Although significant amounts of H$_2$ can co-exist with many
of the metal transitions observed for these models, they still cannot
correctly reproduce the \CaII\ or \CI\ columns.

\subsection{\CII\ fine structure absorption}
Singly ionised carbon (C$^+$) has electronic structure $2s^2 2s^2 2p$
where the outer shell has a configuration $^2$P$^\mathrm{o}_J$, and
thus fine structure splitting occurs between the $J=1/2$ and $J=3/2$
levels. Transitions from these two ground state levels produce
\CII\ and \CIIs\ absorption respectively.

The ratio of C{\sc \,ii}$^*$ to \CII\ column densities has been used
to estimate the star formation rate inferred from the cooling rate in
DLAs at high redshift \citep{Wolfe03}. We find an upper limit on
\CIIs\ from $\lambda 1036$ of $N_\textrm{\CIIs} < 10^{14.5}$\,
\cmm\ and $N_\textrm{\CII} > 10^{14.6}$\, \cmm.  Assuming constant
$n($\CIIs)$/n($\CII) over the entire complex, we find the ratio
$N_\textrm{\CIIs}/N_\textrm{\CII} < 0.8$, consistent with ratios
measured in higher redshift DLAs \citep[for example][]{Srianand05} and
local environments. Following the assumptions described by
\citet{Morris86}, we can estimate the electron density $n_e$ in
\cmmm\ using the expression:
\begin{equation}
\frac{N_\textrm{\CIIs}}{N_\textrm{\CII}} = 3.9 \times 10^{-2}\, n_e
\,[1 + (0.22 \frac{n_p}{n_e})]
\end{equation}
We use $n_p/n_e$ corresponding to the ionised H fraction of $0.70$
from the best-fitting \textsc{Cloudy} ionisation model to find $n_e <
10$\,\cmmm.  Using \NHII\ we can estimate the thickness of the
absorbing cloud as \NHII$/n_e$.  This gives a lower limit on the
cloud size of $3$\,pc. This limit is not necessarily related to the
density or size of the H$_2$ gas, as we argue in the discussion that
most of the \CII\ is due to a warm ionised phase separate from the
H$_2$.

\subsection{Molecular fraction}
\label{s_fhtwo}
The molecular mass fraction is estimated by
\begin{equation}
  f_{\textrm{H}_2} = 2N_{\textrm{H}_2} / (N_{\textrm{\HI}} +
  2N_{\textrm{H}_2}),
\end{equation}
assuming most of the hydrogen associated with the H$_2$ is neutral. In
this case, as for many other QSO absorption systems, it is not clear
how to divide the total \NHI\ measured from the damping wings between
different absorbing components, and in principle each
\NHtwo\ component could have a different \fHtwo\ value. To calculate
\fHtwo\ we use the total \NHtwo\ from both components and
conservatively assume all \NHI\ is associated with the H$_2$, meaning
\fHtwo\ is effectively a lower limit. This gives a molecular fraction
of $\log_{10}$\,\fHtwo\,$= -1.93 \pm 0.36$. As we discuss in the next
section, given the total \NHI, this is an usually high molecular
fraction compared to most other higher redshift systems and sightlines
in the Local Group. Therefore we may be concerned that a different
velocity model to the one we have used permits a much lower \fHtwo. To
calculate a lower limit on \fHtwo\ independent of the velocity model,
we measure the column density of the lowest oscillator strength
transition available for each rotational level ($J=0$, $\lambda
1108.1$; $J=1$, $\lambda 1008.5$; $J=2$, $\lambda 934.1$ and $J=3$,
$\lambda 952.3$) using the AOD method. This gives a lower limit of
\NHtwo $ = 10^{16.5}$~\cmm\ or \fHtwo\,$> 10^{-3}$, again assuming all
of the \NHI\ is associated with the \NHtwo. This is still a high value
relative to local H$_2$ systems with similar \NHI.

\section{Discussion}
\subsection{Physical conditions in the H$_2$ cloud}
\label{s_phys}
\begin{figure*}
\begin{center}
\includegraphics[width=0.97\textwidth]{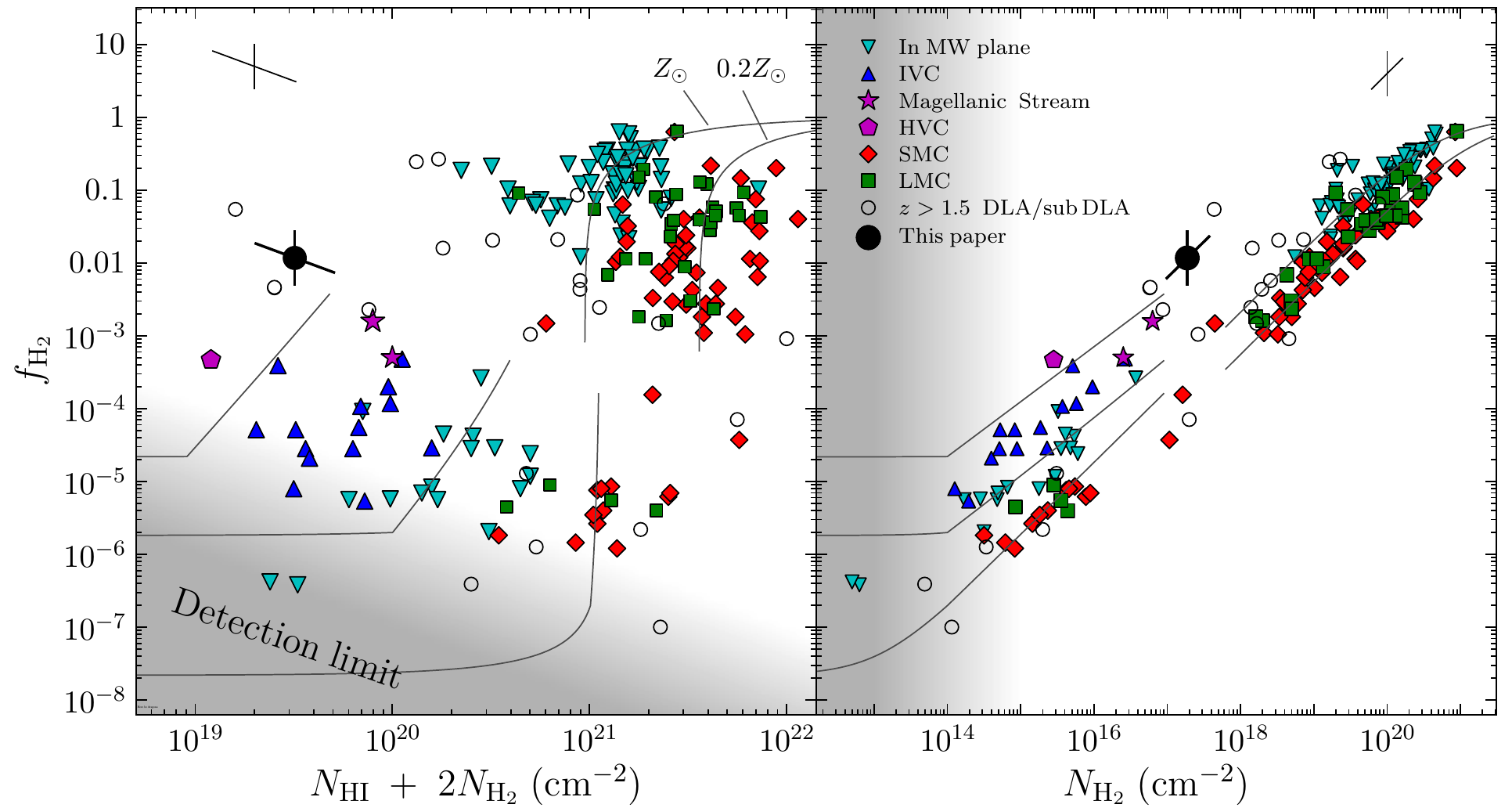}
\caption{\label{f_fntot} The molecular fraction versus the total
  $N_\textrm{H}$. Compared to measurements in the Magellanic clouds
  \citep{Tumlinson02, Welty12}, along the disk of the Milky Way
  \citep{Savage77} and in intermediate- \citep{Richter03b} and
  high-velocity clouds \citep{Richter99}, the $z=0.56$ H$_2$ system
  has an unusually large \fHtwo\ for its total H column. The local
  systems that appear most similar to this absorber are seen along
  sightlines through the Magellanic Stream \citep{Sembach01,
    Richter01b}. Error bars at the corner of each plot show the
  typical uncertainties on \fHtwo, $N_\mathrm{tot}$ and \NHtwo.
  Shading shows the regions in which H$_2$ cannot be detected for a
  limit of \NHtwo\, $\lesssim 10^{14.5}$\, \cmm, a typical threshold
  for the spectra used for the different surveys. The thin lines at
  the top right in each panel show analytic models from Krumholz and
  McKee for two illustrative metallicities. The three thin lines at
  the lower left are the analytic models described in
  Section~\ref{s_phys}.  H$_2$ detections in $z > 1.5$ QSO absorption
  systems (from the compilation by \citealt{Noterdaeme08} with
  additions from \citealt{Petitjean02}, \citealt{Reimers03},
  \citealt{Noterdaeme10}, \citealt{Srianand08, Srianand10,
    Srianand12}, \citealt{Tumlinson10}, \citealt{Jorgenson10} and
  \citealt{Guimaraes12}) are also shown.}
\end{center}
\end{figure*}

To consider this system in the context of other H$_2$ detections in
absorption, we plot \fHtwo\ for local and higher redshift H$_2$
sightlines as a function of the total hydrogen column density and
\NHtwo\ in Figure \ref{f_fntot}. It is apparent that the $z=0.56$
sub-DLA (solid circle) has an unusually high \fHtwo\ given its
\NHI\ compared to sightlines through the plane of the Milky Way (cyan
inverted triangles), or through the Magellanic clouds (green squares
and red diamonds).

Before we discuss the likely origin of the $z=0.56$ sub-DLA, we
examine the physics underlying the \fHtwo\ distribution as a function
of $N_\textrm{H}$ and \NHtwo. The left panel shows a clear bimodality
in the \fHtwo\ $- N_\textrm{H}$ distribution between high
$N_\textrm{H}$, high \fHtwo\ sightlines at the top right, and lower
\fHtwo\ sightlines, generally with much lower $N_\textrm{H}$.  The
right panel shows this is actually a bimodality between \NHtwo\ values
$\lesssim 10^{16}$\,\cmm\ and $\gtrsim 10^{18}$\,\cmm. This can be
understood as the onset of H$_2$ self-shielding against UV
dissociating photons \citep[for example][]{Hirashita05,Gillmon06}. An
analytic approximation from \citet{Draine96} shows that $>97$ per cent
of H$_2$-dissociating photons are blocked by self-shielding once
\NHtwo\ $\sim 10^{16}$\,\cmm. Once H$_2$ becomes
self-shielded\footnote{Dust shielding only becomes important at total
  H columns of $\sim10^{21}$\,\cmm, assuming solar metallicity.}, the
dissociation rate drops and H$_2$ accumulates rapidly to the
formation-destruction equilibrium value predicted by the models by
\citet{McKee10}. These models are shown at the top right in each panel
for two metallicities; solar and $Z=0.2Z_{\odot}$, the metallicity of
the SMC. They were calculated using equations 4, 5, 7 and 8 from
\citet{Kuhlen12} and assume the ISM is in a two phase equilibrium
between a cold neutral medium and a warm neutral medium \citep[for
  example][]{Wolfire95}.  The solar metallicity model reproduces the
mean \fHtwo\ for the Milky Way sightlines through shielded H$_2$
regions reasonably well, although \citet{Welty12} point out that these
models overpredict the $N_\textrm{H}$ at which this transition occurs.

The $N_\textrm{H}$ at which there is sufficient shielding from the UV
field to form large amounts of H$_2$ varies depending on the dust to
gas ratio, the strength of the UV field, and the H$_2$ linewidth, so
the $N_\textrm{H}$ at which the transition from optically thin to
optically thick occurs can change from sightline to sightline. In the
plane of the MW disk, the transition from low to higher \fHtwo\ takes
place around $N_\textrm{H} = 10^{20.5} - 10^{21}\,$\cmm. It occurs at
higher $N_\textrm{H}$ in the LMC and SMC, both because their lower
metallicities \citep[$0.5$ and $0.2Z_{\odot}$ respectively, see the
  appendices from][]{Welty97, Welty99} result in a lower H$_2$
formation rate on grains, and due to an increased UV field compared to
the Milky Way \citep{Tumlinson02}. \citet{Gillmon06} have also shown
the large variation in \fHtwo\ along different sightlines implies that
each sightline intersects a small number of molecule-bearing clouds.

In addition to comparing to the two-phase equilibrium models of McKee
\& Krumholz, we can compare to simple analytic models that apply to
diffuse H$_2$ in the partially shielded regimes.  Following the
Appendix from \citet{Jorgenson10}, the dissociation rate in s$^{-1}$
due to photons with energies corresponding to the Lyman-Werner bands
is given by
\begin{equation}
  R_\textrm{diss} = 1.1\times 10^8\, 4\pi
  J_{\nu}^\textrm{LW}\,S_\textrm{shield},
\end{equation}
where $J_{\nu}^\textrm{LW}$ is the strength of the incident radiation
field in erg~s$^{-1}$cm$^{-2}$Hz$^{-1}$sr$^{-1}$ at 1000\,\AA, and $1
- S_\textrm{shield}$ is the fraction of Lyman-Werner photons processed
by dust or scattered due to \Htwo\ shielding, which can be calculated
using the analytic expressions from \citet{Draine96} and
\citet{Hirashita05}. In formation-dissociation equilibrium, the
molecular fraction can be approximated by
\begin{equation}
  f_{\textrm{H}_2} = 2\, \frac{ \kappa \mathcal{R}
    n_\textrm{\HI}}{R_\textrm{diss}},
\end{equation}
where $\mathcal{R}$ is the formation rate of H$_2$ on dust grains in
cm$^3$\,s$^{-1}$, $n_\textrm{\HI}$ is the \HI\ particle density in
\cmmm\ and $\kappa$ is the dust to gas ratio relative to that in the
solar neighbourhood as defined in Section~\ref{s_kappa}. Note that the
formation rate term used by \citet{Hirashita05}, $R_\mathrm{dust}$, is
equal to $\kappa \mathcal{R}$. Rearranging these expressions, we can
estimate the particle density in the cloud as:
\begin{equation}
  n_\textrm{\HI} = 74\, \cmmm \kappa^{-1}
  \left(\frac{\mathcal{R}}{\mathcal{R}_\mathrm{SN}}\right)^{-1}
  \left(\frac{f_{\textrm{H}_2}}{0.01}\right)\left(\frac{J_{\nu}^\textrm{LW}}{J_{\nu,\,\mathrm{SN}}^\textrm{LW}}
  \right)\left(\frac{S_\mathrm{shield}}{0.01}\right),
\end{equation}
where $J_{\nu,\,\mathrm{SN}}^\textrm{LW}=3.2\times
10^{-20}$\,erg~s$^{-1}$\,cm$^{-2}$\,Hz$^{-1}$\,sr$^{-1}$ and
$\mathcal{R}_\mathrm{SN}=3 \times 10^{-17}$\,cm$^3$\,s$^{-1}$ are
typical values measured in the solar neighbourhood \citep{Habing68,
  Jura74}.

The three curves at the lower left of each panel in
Figure~\ref{f_fntot} show the molecular fractions estimated with
equation (7) for illustrative combinations of $\kappa \mathcal{R}
n_\textrm{\HI} / R_\textrm{diss}$. The upper curve and middle curves
each have $n_\textrm{\HI}=10$\,\cmmm\ with $\kappa=0.04$,
$0.1\mathcal{R}_\mathrm{SN}$, $10 J_{\nu,\,\mathrm{SN}}^\textrm{LW}$
and $\kappa=1$, $0.33\mathcal{R}_\mathrm{SN}$, $2
J_{\nu,\,\mathrm{SN}}^\textrm{LW}$ respectively.  The lower curve has
$\kappa=0.04$, $n_\textrm{\HI}=5$\,\cmmm, $\mathcal{R}_\mathrm{SN}$ and
$0.01 J_{\nu,\,\mathrm{SN}}^\textrm{LW}$. This lower curve has
qualitatively different behaviour from the upper two curves, because
at such low molecular fractions, dust shielding from dissociating
photons becomes important before H$_2$ self-shielding. Therefore the
observed variation in \fHtwo\ can be explained by reasonable
variations in the combination of UV field strength, particle density
and H$_2$ dust formation rate. If the $z=0.56$ system is in H$_2$
formation-dissociation equilibrium, the combination of low \NHI\ and
high molecular fraction suggests that it is either in a weaker UV
field, has an increased H$_2$ formation rate, a higher $\nH$ compared
to the solar neighbourhood, or some combination of these three. We can
use limits on the column densities of the $J=4$ and $5$ levels to put
upper limits on $J_{\nu}^\textrm{LW}$ \citep{Jura75b}. These upper
limits are not very stringent, however, due to weak limits on the
column densities, and constrain $J_{\nu}^\textrm{LW} \lesssim
10^{-18}$\,erg~s$^{-1}$cm$^{-2}$Hz$^{-1}$sr$^{-1}$. This is consistent
with the different UV background values assumed in the \textsc{Cloudy}
models, which range from $\sim 10^{-22}$ for the IGM-like scenario to
$\sim 10^{-20}$\,erg~s$^{-1}$cm$^{-2}$Hz$^{-1}$sr$^{-1}$ for the
ISM-like scenario.

Using \fHtwo\ and \NHtwo\ measured in the $z=0.56$ sub-DLA, the
measured metallicity and equation (8), we find densities of $\sim
1-4$\,\cmmm\ for a UV background incident radiation field, and $\sim
70-480$\,\cmmm\ for a Milky Way ISM-like radiation field. The lower
density range corresponds to cloud thicknesses of $\sim 3 - 10$\,pc,
the high density range to $\sim 0.002 - 0.15$\,pc. The only direct
measurement of the size of a redshifted H$_2$ absorber is by
\citet{Balashev11}, through partial covering of a background QSO broad
line region. They find the region producing H$_2$ is $\sim0.15$~pc and
its surrounding neutral envelope $\sim8$~pc, both of which are
consistent with our size estimates. The upper end of our density range
is consistent with values measured for higher redshift H$_2$ systems
using \CI\ fine structure transitions, but would result in extremely
small cloud sizes.

The low total column density of the $z=0.56$ system suggests that it
does not pass through the ISM of a galaxy.  Returning to
Figure.~\ref{f_fntot}, we see that local systems with similarly low
total $N_\textrm{H}$ and almost as high \fHtwo\ are sightlines through
a high velocity cloud \citep{Richter99} and the Magellanic Stream
\citep{Sembach01,Richter01b}. These clouds have sub-solar
metallicities ($0.3-0.5$ solar), and are most likely tidally stripped
from the Magellanic Clouds (for the Magellanic Stream) or the Milky
Way. \citet{Sembach01} estimate the density of the H$_2$-bearing cloud
they observed in the Magellanic Stream to be $0.3-3$\,\cmmm\ with a
photoionisation rate at least a factor of 10 smaller than the Milky
Way ISM value. The H$_2$ formation timescale for these low densities
is around $1$ Gyr, a large fraction of the estimated lifetime of the
Magellanic Stream \citep[for example][]{Besla10}. Therefore they
favour a scenario where H$_2$ is not formed in place, but has survived
the tidal stripping process and persists due to a combination of
self-shielding and the lower ambient UV field compared to the LMC
ISM. Such a scenario could also be responsible for the $z=0.56$
absorber.

\subsection{Comparison to higher-redshift H$_2$ absorbers}

Unlike the local sightlines, there is no clear bimodality in the
\fHtwo\ -- $N_\textrm{H}$ distribution for higher-$z$ H$_2$ systems.
This could be due to each higher-$z$ absorber being comprised of
several clouds, or to a much wider range of incident UV and H$_2$
formation rates, both of which may smooth away an underlying
distribution.

The three high-$z$ systems with \fHtwo\ and $N_\textrm{H}$ most
similar to this system are those described by \citet{Petitjean02} (at
$z=1.973$ towards Q0013-0029 with $\NHI \le 10^{19.4}$~\cmm,
$\fHtwo=10^{-2.63}$), \citet{Reimers03} ($z=1.51$ towards
HE0515$-$4414 with $\NHI = 10^{19.88}\,$\cmm, $\fHtwo=10^{-2.64}$),
and \citet{Tumlinson10} and \citet{Milutinovic10} ($z=2.059$ towards
Q2123$-$0500 with $\NHI = 10^{19.18}\,$\cmm, $\fHtwo=10^{-1.54}$). The
$z=1.973$ system is a sub-DLA component that is highly depleted to the
same extent as is observed for cool gas in the Milky Way. It has a
solar metallicity and the gas pressure is even higher than is
typically measured in Milky Way ISM. The $z=1.51$ system has a
metallicity of 0.3 solar, and dust to gas ratio of $0.89\pm 0.19$
relative to solar. It also shows evidence of a higher
photodissociation rate than is seen locally. The final sub-DLA at
$z=2.059$ has a metallicity of 0.5 solar, and HD absorption is
observed in addition to H$_2$. It exhibits a multi-phase medium of
cold and warm gas, similar to the system we have presented is this
paper. Unfortunately none of these absorbers have associated imaging
to suggest a typical impact parameter of any nearby galaxy producing
the absorption.

Therefore, the three higher redshift systems showing a similarly high
\fHtwo\ and low \NHI\ tend to have larger metallicities and dust to
gas ratios than the $z=0.56$ absorber. However, it is possible that
the components producing H$_2$ in the $z=0.56$ system have a higher
metallicity and dust-to-gas ratio than that averaged over the whole
absorber.

\subsection{Connection to galaxies}
\label{s_gal}
\begin{figure}
\begin{center}
\includegraphics[width=0.47\textwidth]{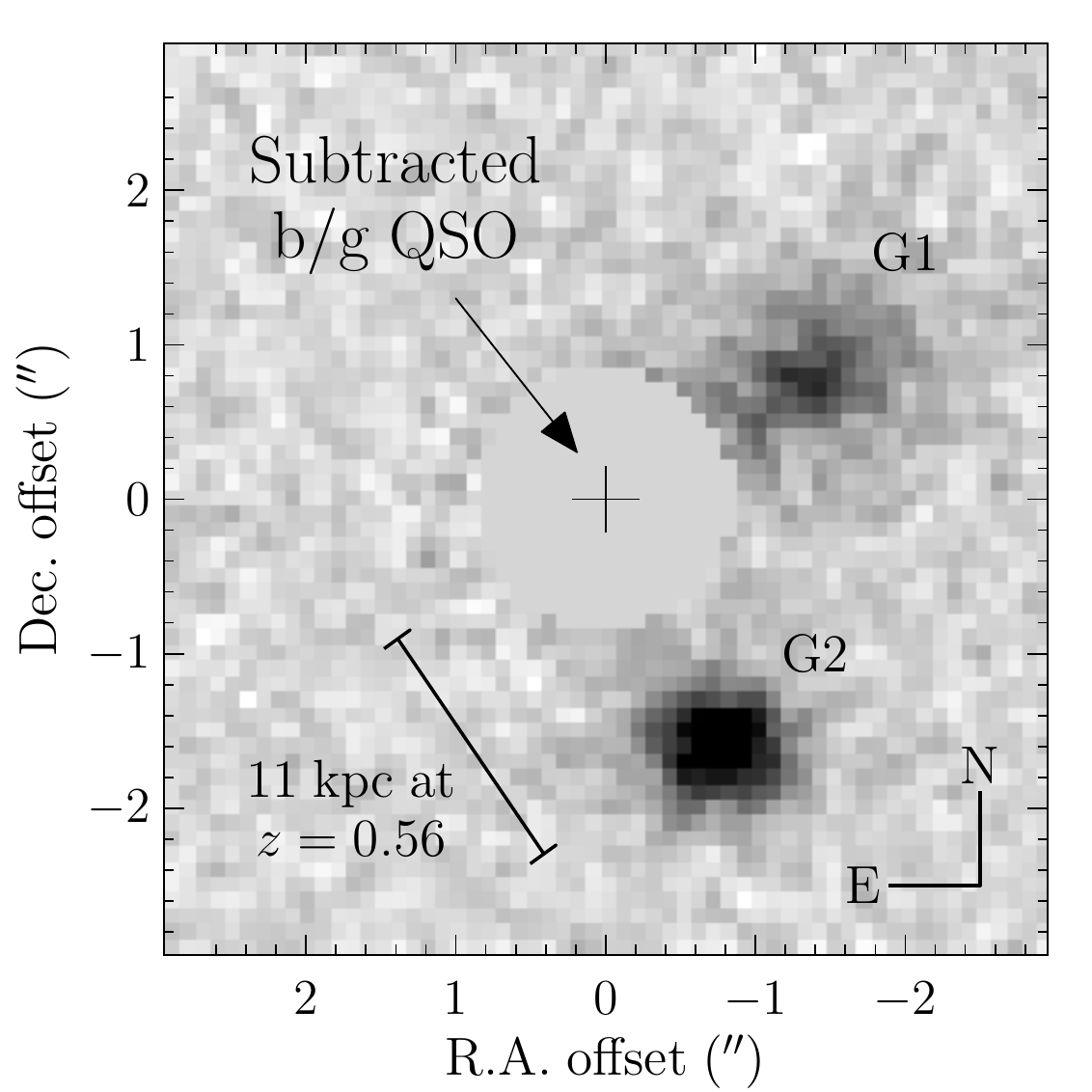}
\caption{\label{f_im} A K-band image of the QSO showing two nearby
  galaxy candidates, G1 and G2. The QSO image has been subtracted
  using the point spread function of a nearby star. The region near
  the centre of the QSO still shows significant residuals and has been
  greyed out. If G1 and G2 are at the redshift of the QSO then they
  have impact parameters of $10$~kpc and $11$~kpc, and luminosities of
  $0.7L^*$ and $2L^*$ respectively.}
\end{center}
\end{figure}

The $K$ band imaging around Q~0107$-$0232 has a seeing FWHM of
0.8\arcsec, and shows two possible galaxy candidates less than
1.2\arcsec\ from the QSO sightline. Figure \ref{f_im} shows a
$5$\arcsec$\times5$\arcsec\ region centred on the QSO. The QSO image
has been subtracted using the point spread function of a nearby
star. The two galaxy candidates are seen to the North-West (G1) and
South-West (G2). Assuming they are at the redshift of the absorber,
they have luminosities of $0.7L^*$ (G1) and $2L^*$ (G2), and impact
parameters of $10$~kpc (G1) and $11$~kpc (G2), both smaller than the
median impact parameter of $33$~kpc for galaxies associated with
sub-DLAs found by \citet{Rao11}. Therefore it is likely that at least
one is associated with the absorber, on scales typical of the
separations between the Milky Way and high-velocity clouds (10-60~kpc,
see \citealt{Putman12} and the references therein).

\subsection{Three different gas phases in the sub-DLA}
\label{s_threephase}
Figures~\ref{f_cloudy1} \& \ref{f_cloudy2} show the total hydrogen
particle density for the majority of metals observed in this system
corresponding to the ionisation parameter, assuming the normalisations
of the incident radiation fields are correct. The most likely model
corresponds to hydrogen densities from $10^{-3}$ to
$10^{-2}$~\cmmm. Even assuming a factor of ten uncertainty in the
radiation field strength, this is much lower than the typical
densities where H$_2$ is seen in both our galaxy and in other
H$_2$-bearing DLAs ($\nH = 10-100$~\cmmm). This is confirmed by our
\textsc{Cloudy} modelling, which shows that there is no single cloud
model that can simultaneously reproduce both the \CI\ and H$_2$ column
densities, in addition to those of the other low-ionisation metal
transitions. Therefore the gas traced by most of the metal absorption
is probably in a different environment to that in which the H$_2$
resides. This is also likely the cause of the excess $N_\textrm{\CI}$
over that predicted by the \textsc{Cloudy} models. \CI\ is often seen
in dense components showing H$_2$ \citep[for example][]{Srianand05}
and can have extremely narrow linewidths corresponding to temperatures
of $\sim 100$\,K \citep{Jorgenson09,Carswell11}, indicating it occurs
in the same environment as H$_2$. Thus most of the \CI\ and some
\CaII\ may be from a high density region co-spatial with the H$_2$.

As discussed in Section~\ref{s_cloudy}, the presence of \OVI\ is
unlikely to be explained by photoionisation by a hard UV field. At the
metallicity of the absorber ($\sim 0.1$ solar), significant \OVI\ is
only produced via collisional ionisation for temperatures larger than
$10^5$\ K, even in non-equilibrium cases \citep{Gnat07}. Thus it is
likely a hotter medium than that producing the H$_2$ and metal lines
is also present.

We conclude that the absorption is due to gas in three phases: a
photoionised medium at $\sim 10^4$\,K in which most of the metal
transitions we see are produced, a cold neutral medium at $\sim
100$\,K where the H$_2$ and \CI\ absorption occurs, and a hotter phase
where \OVI\ is produced. The \HI\ column is likely split between the
two cooler phases. A similar multi-phase environment is also seen in
other higher redshift sub-DLAs that show molecular absorption
\citep{Milutinovic10}.\footnote{Milutinovic et al. did not report an
  \OVI\ detection, but as this system was at a higher redshift, it may
  have been heavily blended with \lya\ forest absorption.}

The Magellanic Stream and many other HVCs comprise $10^4$~K ionised
gas that is seen in H$\alpha$ emission, T~$>10^5$~K hot gas producing
\OVI\ absorption, and they can also contain cold neutral gas with
H$_2$ \citep{Sembach03, Fox10}. Taken together, the existence of these
three phases, the high molecular fraction with a low total column
density, and the proximity of a possible $\sim L^*$ galaxy suggest the
$z=0.56$ absorber is due to a tidally stripped feature analogous to
the Magellanic Stream.

\subsection{Incidence rate of H$_2$ in low redshift sub-DLAs}
Due to the need for bright targets observable with space-based UV
spectroscopy and their low incidence rate, very few DLAs and sub-DLAs
have been found at low redshift. Until recently only $\sim 10$ DLAs at
redshifts $<1$ were known, and only a handful of these have coverage
of H$_2$ Lyman-Werner bands. With the availability of COS, the number
of such systems is being increased dramatically, and due to its far UV
wavelength coverage the presence of H$_2$ can be easily detected.

\citet{Battisti12} present a sample of $2$ DLAs and $6$ sub-DLAs at $z
< 0.35$, serendipitously discovered along sightlines as part of a
large COS program. Like the sub-DLA presented here, they were not
pre-selected by the strength of their metal lines or other properties
that might influence the likelihood of detecting
molecules. Interestingly, they also discovered a sub-damped system
with H$_2$ absorption at $z=0.2477$. Taking this sample together with
the system in this paper and assuming binomial statistics, we find the
expected incidence rate of DLAs and sub-DLAs showing molecular
hydrogen at \NHtwo\,$\gtrsim10^{14}$\,\cmm\ at low redshift to be $2 /
9 = 22$ per cent (with a $95$ per cent confidence level lower limit of
$4$ per cent), rising to $33$ ($5$) per cent if we consider only the
sub-damped systems with \NHI\,$< 10^{20}$\,\cmm.  This is a
surprisingly large fraction given that sub-DLAs are often found to be
highly ionised absorbers with $\lesssim 10$ per cent of their hydrogen
in the form of \HI.

If we think that the absorption cross section for H$_2$ is dominated
by cold gas associated with Local Group-type systems (the Magellanic
Stream for example), then this may be consistent with this high
incidence rate. \citet{Richter12} shows that one can explain 30-100
per cent of the observed incidence rate of systems with \NHI\,$>
10^{17.5}$\, \cmm\ as intermediate- and high-velocity clouds
distributed around galaxies with \HI\ masses between $10^{8.5}$ and
$10^{10}$ M$_{\sun}$ in a similar way as is seen around M31 and the
Milky Way. As discussed in the previous section, some HVCs also show
relatively high molecular fractions, and in terms of \NHI\ and
\fHtwo\ HVCs are the local systems most analogous to the system
analysed in this paper.

It would be interesting to perform a systematic search for H$_2$ in
further $10^{19}$\,\cmm\ $< $~\NHI~$ < 10^{20.3}$\,\cmm\ sub-DLAs at
both high and low redshifts that have metal absorption consistent with
a cool, dusty environment. Sub-DLAs tend to have both higher
metallicities and larger velocity widths than DLAs, and H$_2$ is more
likely to be found in DLAs with both these characteristics
\citep{Noterdaeme08}.

\subsection{Evolution in $\bmath{\fHtwo}$}
\begin{figure}
\begin{center}
\includegraphics[width=0.4\textwidth]{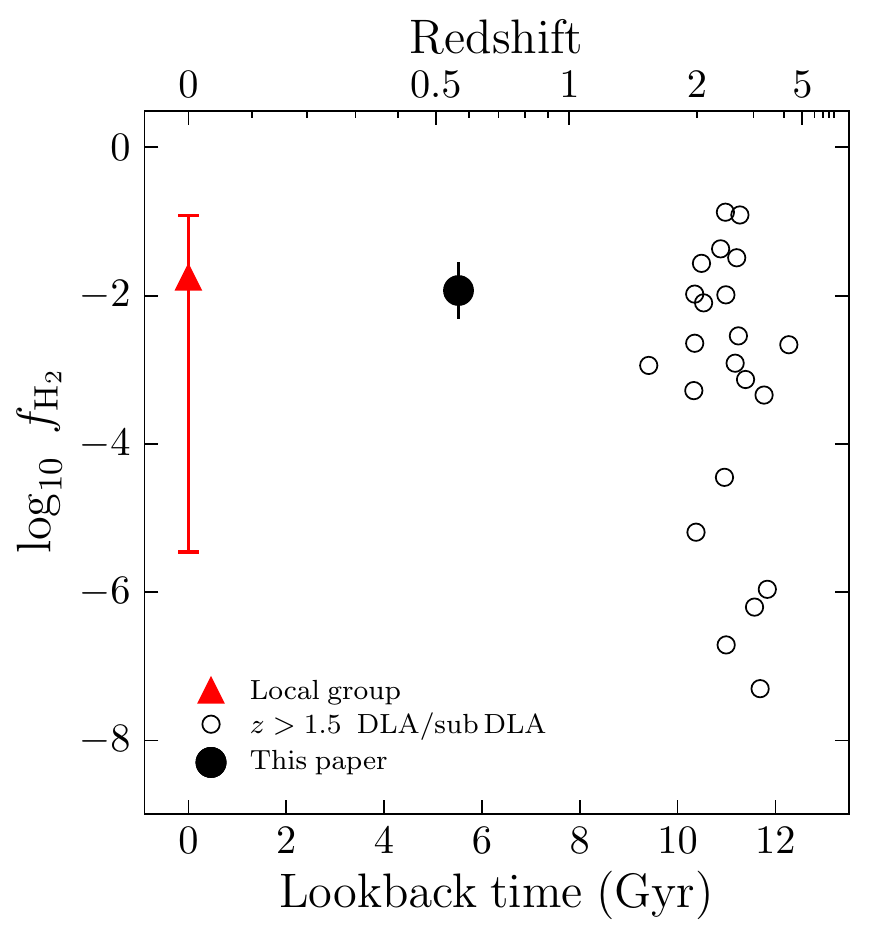}
\caption{\label{f_fz} The molecular mass fraction as a function of
  lookback time for sites where H$_2$ absorption has been
  detected. The Local Group point is a median for all values from the
  plane of the Milky Way \citep{Savage77}, the LMC and SMC
  \citep{Welty12}. The error bars show the 10th and 90th percentile
  level. There is no evidence for evolution in \fHtwo, but the number
  of measurements at $z > 0$ is small.}
\end{center}
\end{figure}
We plot the \fHtwo\ values as a function of cosmic time in
Figure~\ref{f_fz}. There is no evidence for evolution in \fHtwo,
though more measurements are needed, particularly at intermediate
redshifts, given the large scatter in \fHtwo\ seen along both local
sightlines in the Milky Way halo and in higher $z$ DLAs.

\section{Summary}
We have analysed a sub-damped \lya\ system with \NHI $=10^{19.5 \pm
  0.2}$\,\cmm\ at $z=0.56$ that shows associated molecular hydrogen
absorption in the Lyman and Werner bands. Using velocity components
determined from a high resolution spectrum covering metal transitions
falling in the optical, we fit a two-component model to the H$_2$
absorption and find a lower limit to the molecular fraction of
$\log_{10} \fHtwo = -1.93 \pm 0.36$, and a lower limit independent of
the assumed velocity structure of \fHtwo\,$>10^{-3}$.  This is higher
than other sightlines with similar \NHI\ where H$_2$ has been measured
in the Milky Way halo. We find a metallicity for the cloud $\log_{10}
Z = -0.72\pm 0.32$, or $0.19^{+0.21}_{-0.10}$ solar. The dust-to-gas
ratio relative to the solar neighbourhood is $\log_{10} \kappa <
-0.44$, or $\kappa < 0.36$.

We modeled the observed transitions using \textsc{Cloudy} and were
unable to find a single solution that can simultaneously reproduce all
the observed transitions. However, a model for the absorber of a
$10^4$\,K cloud illuminated by a radiation field dominated by the UV
background can broadly reproduce all the observed column densities
apart from those of H$_2$, \CI, and \OVI. We conclude that there are
three phases in the absorber; a $T\sim 100$\,K phase where the
\CI\ and H$_2$ arise, a $T\sim 10^4$\,K phase where the
low-ionisation metal absorption occurs, and a hotter, collisionally
ionised phase associated with \OVI.

Using simple models of H$_2$ formation-dissociation equilibrium, we
calculate densities for the H$_2$ absorbing region from $\sim
1-4$\,\cmmm\ to $\sim 70-480$\,\cmmm, depending on the incident
strength of the radiation field. The lower density range corresponds
to cloud thicknesses of $\sim 3 - 10$\,pc, the high density range to
$\sim 0.002 - 0.15$\,pc. Given the \NHI, the presence of a three phase
medium, the molecular fraction, metallicity and two galaxy candidates
near the QSO sightline with impact parameters of $\sim 10$\,kpc, we
conclude this system may be a tidally stripped feature similar to the
Magellanic Stream.

Finally, we remark that of the seven sub-DLAs observed at $z < 0.7$
for which there is the possibility to detect \NHtwo\,$\gtrsim
10^{14.5}$\,\cmm, two H$_2$ detections were found. A survey for H$_2$
in low-redshift sub-damped systems could be a fruitful way to measure
the physical conditions giving rise to these absorbers.

\vspace{0.5cm}

We acknowledge helpful correspondence with Jason Tumlinson and thank
Andrew Fox, Joe Hennawi, Mark Krumholz, Kate Rubin, Karin Sandstrom
and Todd Tripp for illuminating conversations, and Dan Welty for
comments on an earlier version of this paper. Gabor Worseck kindly
helped us to obtain the HIRES spectrum of Q~0107$-$0232. We thank the
anonymous referee for comments that helped improve the paper.

Some of the data presented here were taken at the W.M. Keck
Observatory, which is operated as a scientific partnership among the
California Institute of Technology, the University of California and
the National Aeronautics and Space Administration. The Observatory was
made possible by the generous financial support of the W.M. Keck
Foundation. This analysis made use of observations from the NASA/ESA
Hubble Space Telescope, obtained at the Space Telescope Science
Institute, which is operated by the Association of Universities for
Research in Astronomy, Inc., under NASA contract NAS 5-26555 (program
11585) and of observations collected at the European Organisation for
Astronomical Research in the Southern Hemisphere, Chile (program
383.A-0402).

The authors wish to recognize and acknowledge the significant cultural
role and reverence that the summit of Mauna Kea has always had within
the indigenous Hawaiian community. We are most fortunate to have the
opportunity to conduct observations from this mountain.

N.T. acknowledges grant support by CONICYT, Chile (PFCHA/{\it
  Doctorado al Extranjero 1$^{\rm a}$ Convocatoria}, 72090883). Most
of the programs particular to this analysis were written using the
NumPy and SciPy packages (\url{http://www.scipy.org}), and plots were
produced using Matplotlib
\citep[][\url{http://www.matplotlib.sourceforge.net}]{Hunter07}.

\footnotesize{
  \bibliographystyle{./mn2e}

}

\appendix
\section{Correcting for wavelength shifts in the COS exposures.}
\label{a_wshifts}
When combining the COS exposures, we found there were shifts of $\sim
20$~\kms\ in the wavelength solution between exposures taken using
different central wavelength settings. Table~\ref{t_wac_shifts} gives
the shifts that must be applied to bring our G160M exposures with
central wavelength settings 1589 and 1627\,\AA\ to a common wavelength
scale. Table~\ref{t_H2_shifts} gives the further $\sim
10$~\kms\ shifts that were required to give an internally consistent
wavelength solution based on the expected positions of H$_2$
absorption features.
\begin{table}
\begin{center}
\begin{tabular}{cccccc}
\hline
& Valid & Apply to  & & \\ 
Segment & $\lambda$ range (\AA) & $\lambda$ setting & $\alpha$ & $\beta$ \\
\hline
FUVA & $1600$ -- $1800$ & 1627 & $ 1.233 \times 10^{-3}$ & $-$2.115 \\
FUVB & $1425$ -- $1600$ & 1589 & $-0.952 \times 10^{-3}$ &  1.468 \\
\hline
\end{tabular}
\caption{\label{t_wac_shifts} Wavelength shifts required to bring COS
  G160M exposures with central wavelength settings 1589 and
  1627\,\AA\ on to a common wavelength scale. The shift $\Delta
  \lambda$ that must be added to a wavelength $\lambda$ is given by
  $\Delta \lambda = \alpha \lambda + \beta$. The FUVA shift is applied
  to the 1627 central wavelength setting, and the FUVB shift is
  applied to the 1589 setting.}
\end{center}
\end{table}
\begin{table}
\begin{center}
\begin{tabular}{cccc}
\hline
$\lambda$ (\AA) & $\Delta \lambda$ (\AA)& $\lambda$ & $\Delta \lambda$ \\ 
\hline
$1425$ & $-0.1009$ & $1625$ & $-0.0062$\\
$1450$ & $-0.0848$ & $1650$ & $-0.0033$\\
$1475$ & $-0.0679$ & $1675$ & $-0.0011$\\
$1500$ & $-0.0514$ & $1700$ & $0.0009 $\\
$1525$ & $-0.0365$ & $1725$ & $0.0029 $\\
$1550$ & $-0.0245$ & $1750$ & $0.0052 $\\
$1575$ & $-0.0160$ & $1775$ & $0.0080 $\\
$1600$ & $-0.0101$ & $1800$ & $0.0116 $\\
\hline
\end{tabular}
\caption{\label{t_H2_shifts} Wavelength shifts inferred from the
  H$_2$ absorption as function of wavelength. $\Delta \lambda$ should
  be added at each wavelength $\lambda$ to correct the wavelength
  scale.}
\end{center}
\end{table}
\end{document}